\documentclass[12pt]{article}
\usepackage{a4wide}
\usepackage{epsfig}

\newcommand{\Li}{\mathop{\rm Li}\nolimits}
\newcommand\pfrac[2]{\left(\frac{#1}{#2}\right)}
\newcommand{\GeV}{\mathop{\rm GeV}\nolimits}

\begin{document}

\thispagestyle{empty}

\begin{center}
{\Large\bf $O(\alpha_s)$ perturbative and nonperturbative corrections to\\[7pt]
  polarized semileptonic $\Lambda_b$ decay distributions}\\[7mm]
{\large M. Fischer$^1$ and S.~Groote$^2$}\\[7mm]
$^1$ Institut f\"ur Physik, Johannes-Gutenberg-Universit\"at,
  D-55099 Mainz, Germany\\[.2cm]
$^2$ Loodus- ja t\"appisteaduste valdkond, F\"u\"usika Instituut,\\[.2cm]
  Tartu \"Ulikool, W.~Ostwaldi 1, EE-50411 Tartu, Estonia\\[7pt]
\end{center}

\vspace{7mm}

\begin{abstract}
In this paper we calculate first order radiative QCD corrections to the
decay process $b\to cW^-(\to\ell^-\bar\nu_\ell)$ of a polarized bottom quark.
Taking into account both the bottom and charm quark masses, the analytical
expressions given in this paper allow for a detailed analysis of the
differential decay rate of the polarized quark in dependence on the polar and
azimuthal angles of the lepton pair and the angle of the polarization vector
of the quark relative to the momentum direction of the $W$ boson. The results
are given for different polarization states of the charged lepton. In
addition, we calculated nonperturbative corrections and estimated their
contribution.
\end{abstract}

\newpage

\section{Introduction}
The top quark as the heaviest known fermion is an ideal test particle, as it
decays before the hadronization process can take place. Therefore, the quantum
state of this particle is transferred mostly unperturbed to the decay products.
While top quarks and antiquarks produced in pairs are mostly
unpolarized~\cite{CMS:2013roq,ATLAS:2016bac}, top quarks produced in single
top production are found to be highly polarized due to parity
violation~\cite{Mahlon:1998uv,Espriu:2002wx,Rindani:2015dom,CMS:2015cyp,%
ATLAS:2017ygi}. As the top quark decays with more than $99\%$ probability
via the channel $t\to bW^+$, the question arises whether the polarization of
the top quark gained in the production process is passed on to the decay
products, i.e.\ the $W$ boson and the bottom quark.

The Standard Model (SM) of elementary particle physics describes the
polarization states of the $W$ boson, measured by the subsequent decay of the
$W$ boson into e.g.\ a pair of lepton and antineutrino. Any deviation from
this prediction opens a window for Beyond Standard Model (BSM) physics.
Therefore, QCD and electroweak radiative corrections for the process $t\to bW$
have been calculated in order to distinguish BSM effects from perturbative and
nonperturbative SM contributions (cf.\ e.g.\ Refs.~\cite{Drobnak:2010ej,%
Rindani:2011pk,Ajaltouni:2019log}). In a couple of publications, we have given
our contribution to this analysis, calculating electroweak corrections to
next-to-leading (NLO) order~\cite{Do:2002ky,Groote:2017vux} and QCD corrections
to NLO~\cite{Fischer:1998gsa,Fischer:2001gp} and next-to-next-to leading (NNLO)
order~\cite{Czarnecki:2018vwh}.

Another channel for the production of polarized bottom quarks is the decay of
the $Z$ boson. If it is possible to separate the $Z$ boson from the photon
with which it is mixed e.g.\ on $e^+e^-$ annihilations, the decay of the $Z$
boson to a heavy quark--antiquark pair also shows a significant degree of
longitudinal polarization which amount to $\langle P_b\rangle=-0.94$ for
bottom quarks and $\langle P_c\rangle=-0.68$ for charm
quarks~\cite{Kuhn:1992ndv}. In a sequence of papers the possibility to measure
the polarization of bottom (and charm) quarks at ATLAS and CMS has been
analyzed~\cite{Galanti:2015pqa,Kats:2015cna,Kats:2015zth,Kats:2017wuh}. The
authors come to an affirmative answer, though the quarks are observed only as
a jet of hadrons. As noted already in Ref.~\cite{Close:1991sp}, at least in
the heavy quark limit the polarization transfer from a heavy quark to the
heavy hadron approaches 100\%, i.e.\ a hadron like $B$ or $\Lambda_b$ can be
considered to carry the same polarization as the bottom quark. For a less
idealistic situation, Falk and Peskin have estimated the reduction factor to
be of the order of $75\%$~\cite{Falk:1993rf}.

In this paper we analyze the polarization of the bottom quark by considering
the energy and angular distribution of the leptons in the cascade process
$b\to cW^-(\to\ell^-\bar\nu_\ell)$, taking into account both QCD corrections
and the exact masses of both bottom and charm quark. In Born approximation,
the differential rate is given by~\cite{Tsai:1978rm,Czarnecki:1993gt}
\begin{eqnarray}
\frac{d\Gamma_b^{(0)}}{dx\,d\cos\theta}&=&\frac{G_F^2m_b^5}{32\pi^5}
  \frac{x^2(1-x-y^2)^2}{6(1-x)^2}\times\strut\nonumber\\&&\strut
  \left[3-2x+y^2+\frac{2y^2}{1-x}
  +S\cos\theta\left(1-2x+y^2-\frac{2y^2}{1-x}\right)\right],\qquad
\end{eqnarray}
where $y=m_c/m_b$ is the scaled charm quark mass and $x=2E_\ell/m_b$ is the
scaled energy of the charged lepton, ranging between $0$ and $1-y^2$. $S=1$
corresponds to a fully polarized bottom quark, while $S=0$ gives the result
for an unpolarized bottom quark.

In Ref.~\cite{Czarnecki:1993gt} it is emphasized that while QCD corrections
modify significantly the decay rate and, therefore, the life time of the
heavy quarks~\cite{Cabibbo:1978sw}, this is not true for the energy
distribution of the charged lepton~\cite{Ali:1979is,Cabibbo:1979jg,%
Corbo:1982ad,Altarelli:1982kh,Corbo:1982ah,Jezabek:1988ja}. However, the
parts dependent and independent of the polarization may be affected
differently. Therefore, the calculation of QCD corrections turns out to be
necessary.

\begin{figure}[htb]
 \begin{center}
   \includegraphics[width=8cm]{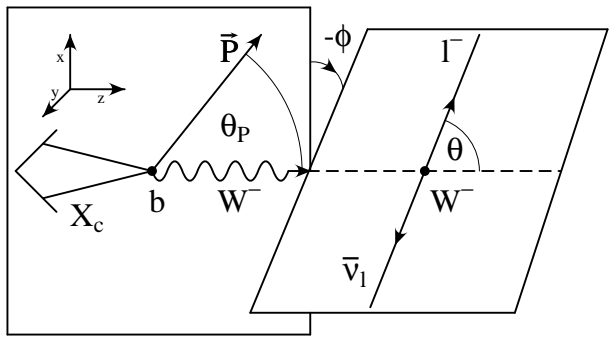}
 \end{center}
\caption{\label{kinematics}Definition of the angles $\theta_P$, $\theta$, and
  $\phi$ in the decay process $b^\uparrow\rightarrow c+\ell^-+\bar\nu_\ell$.
  $\theta_P$ is the polar angle between the direction of the $W$ boson 
  momentum in positive $z$ direction and the polarization vector of the bottom
  quark in the rest frame of the bottom quark which define the hadron plane.
  $\theta$ is the polar angle between the $z$ direction and the lepton
  momentum in the rest frame of the $W$ boson which define the lepton plane.
  The azimuthal angle between hadron and lepton plane is denoted by $\phi$.}
\end{figure}

QCD corrections to the energy distribution have been calculated in
Refs.~\cite{Ali:1979is,Corbo:1982ad,Corbo:1982ah} which agree with numerical
results translated from top quark decays~\cite{Jezabek:1987nf}. QCD
corrections to the joint energy and angular distribution have been calculated
in Ref.~\cite{Jezabek:1987nf}. In Ref.~\cite{Czarnecki:1993gt} QCD corrections
to the angular distribution has been calculated for the charm quark decay
where the mass of the quark in the decay channel has been neglected. This,
however, is not appropriate for the decay of a bottom quark into a charm
quark. For this reason, in this publication we calculate QCD corrections for
both energy and angular distributions, taking into account the masses of
both quarks. In addition to the dependence on the polar angle $\theta$
between the direction of the $W$ boson and the charged lepton, boosted back to
the rest frame of the $W$ boson, we also take into account the azimuthal angle
$\phi$ between the plane spanned by these and the plane spanned by the $W$
momentum and the polarization vector of the bottom quark with a relative angle
$\theta_P$ in the rest frame of the bottom quark. This kinematics is shown in
Figure~\ref{kinematics}.

The paper is organized as follows. Starting from the hadron and lepton tensor
given separately, in Sec.~2 we calculate the differential decay rate. The
hadron tensor, responsible for the inclusive decay $b^\uparrow\to cW^-$, is
written in terms of five unpolarized and nine polarized covariants. In Sec.~3
we calculate the corresponding coefficients, called invariant structure
functions, and relate them to another set of structure functions more
appropriate for the angular distribution and obtained from the hadron tensor
by applying covariant helicity projection operators. In Secs.~4 and~5 we
calculate the first order QCD three graph and one loop corrections. In Sec.~6
we integrate the former over the energy of the intermediate $W$ boson and add
up both contributions. In Sec.~7 we give nonperturbative corrections. In
Appendix~A we give explicit results for the integrated helicity rates.

\section{Differential decay rate}
The differential decay rate for the process
\begin{equation}
b^\uparrow(p_b)\rightarrow\ell^-(p_\ell)+\bar\nu_\ell(p_\nu)+X_c(p_c)
\end{equation}
is given by the contraction of the hadron tensor with the lepton tensor,
\begin{equation}
\frac{d\Gamma}{dq^2dE_\ell}=\frac{G_F^2|V_{bc}|^2}{(2\pi)^3m_b}
  W_{\mu\nu}(p_b,s_b,q)L^{\mu\nu}(p_\ell,p_\nu),
\end{equation}
where $q=p_\ell+p_\nu$ is the four momentum of the $W$ boson and $E_\ell$ is
the charged lepton energy in the rest frame of the decaying bottom quark. In
addition to the dependencies on momenta, we consider the dependence on the
spin $s_b$ of the bottom quark. For massless neutrinos the lepton tensor is
given by
\begin{equation}
L^{\mu\nu}=p_\ell^\mu p_\nu^\nu+p_\ell^\nu p_\nu^\mu
  -p_\ell\cdot p_\nu g^{\mu\nu}
  +i\epsilon^{\mu\nu\rho\sigma}p_{\ell,\rho}p_{\nu,\sigma}
\end{equation}
where we use the convention $\epsilon_{0123}=+1$. Based on Lorentz covariance
and CP symmetry, the hadron tensor can be written in terms of five unpolarized
and nine polarized invariant structure functions,
\begin{eqnarray}
W_{\mu\nu}(p_b,s_b,q)&=&-g_{\mu\nu}W_1(x)+v_\mu v_\nu W_2(x)
  -i\epsilon_{\mu\nu\rho\sigma}v^\rho x^\sigma W_3(x)
  \nonumber\\[7pt]&&\strut
  +x_\mu x_\nu W_4(x)+(v_\mu x_\nu+x_\mu v_\nu)W_5(x)
  \nonumber\\[7pt]&&\strut
  -(x\cdot s_b)\Big[-g_{\mu\nu}W^P_1(x)+ v_\mu v_\nu W^P_2(x)
  -i\epsilon_{\mu\nu\rho\sigma}v^\rho x^\sigma W^P_3(x)
  \nonumber\\[7pt]&&\strut\qquad
  +x_\mu x_\nu W^P_4(x)+(v_\mu x_\nu+x_\mu v_\nu)W^P_5(x)\Big]
  \nonumber\\[7pt]&&\strut
  +(s_{b,\mu}v_\nu+s_{b,\nu}v_\mu)W^P_6(x)
  +(s_{b,\mu}x_\nu+s_{b,\nu}x_\mu)W^P_7(x)\nonumber\\[7pt]&&\strut
  +i\epsilon_{\mu\nu\rho\sigma}v^\rho s_b^\sigma W^P_8(x)
  +i\epsilon_{\mu\nu\rho\sigma}x^\rho s_b^\sigma W^P_9(x),
\end{eqnarray}
where $v=p_b/m_b$ and $x=q/m_b$ are momenta normalized to the mass of the
bottom quark.\footnote{Note that while $x$ is defined as scalar in the
Introduction, at this point we use the same symbol for a four-vector. Still,
this should not cause problems, as $p_\nu=q-p_\ell$ for vanishing lepton masses
leads to $q^2=m_\nu^2-m_\ell^2+2qp_\ell=2E_\ell\sqrt{q^2}$, $\sqrt{q^2}=2E_\ell$
and, therefore, $x_\mu x^\mu=q^2/m_b^2=4E_\ell^2/m_b^2=x^2$.} Note that instead
of the traditional notation we use $W^P_i=G_i$ for the polarized structure
functions. This helps to keep track of the different structure functions.

In the rest frame of the $b$ quark and with the $z$ axis along the momentum of
the $W$ boson, the parametrization of the momentum and polarization four
vectors explicitly reads
\begin{eqnarray}
p_b=(m_b,0,0,0),&&
s_b=(0,\sin\theta_P,0,\cos\theta_P),
  \nonumber\\[7pt]
p_\ell=(E_\ell,-|\vec p_\ell|\sin\theta\cos\phi,
  -|\vec p_\ell|\sin\theta\sin\phi,|\vec p_\ell|\cos\theta),&&
q=(q_0,0,0,|\vec q\,|),\nonumber\\[7pt]
p_c+p_G=p_b-q,&&
p_\nu=q-p_\ell.
\end{eqnarray}
Inserting lepton and hadron tensors, using
$\epsilon_{\mu\nu\rho\sigma}\epsilon^{\mu\nu\tau\omega}=-2
(\delta_\rho^\tau\delta_\sigma^\omega-\delta_\rho^\omega\delta_\sigma^\tau)$
and
\begin{equation}
2p_\ell q=q^2+p_\ell^2-p_\nu^2=q^2+m_\ell^2,\qquad
2p_\nu q=q^2-p_\ell^2+p_\nu^2=q^2-m_\ell^2,
\end{equation}
one obtains for the differential decay rate
\begin{eqnarray}
\lefteqn{\frac{d\Gamma}{dq^2dE_\ell}
  \ =\ \frac{G_F^2|V_{b c}|^2}{(2\pi)^3}\Bigg\{
  (q^2-m_\ell^2)W_1-\frac12(4E_\ell^2-4E_\ell q_0+q^2-m_\ell^2)W_2}
  \nonumber\\&&\strut
  -\frac1{m_b}\left(2E_\ell q^2-q_0(q^2+m_\ell^2)\right)W_3
  +\frac{m_\ell^2}{2m_b^2}(q^2-m_\ell^2)W_4
  -\frac{2m_\ell^2}{m_b}(E_\ell-q_0)W_5\nonumber\\&&\strut
  +\frac{|\vec q\,|}{m_b}\cos\theta_P\Bigg[(q^2-m_\ell^2)W^P_1
  -\frac12(4E_\ell^2-4E_\ell q_0+q^2-m_\ell^2)W^P_2
  \nonumber\\&&\strut\qquad
  -\frac1{m_b}\left(2E_\ell q^2-q_0(q^2+m_\ell^2)\right)W^P_3
  +\frac{m_\ell^2}{2m_b^2}(q^2-m_\ell^2)W^P_4
  -\frac{2m_\ell^2}{m_b}(E_\ell-q_0)W^P_5\Bigg]\nonumber\\&&\strut
  -2\left(E_\ell|\vec q\,|\cos\theta_P-(q_0-2E_\ell)|\vec p_\ell|
  (\cos\theta_P\cos\theta-\sin\theta_P\sin\theta\cos\phi)\right)W^P_6
  \nonumber\\&&\strut
  -2\frac{m_\ell^2}{m_b}\left(|\vec q\,|\cos\theta_P-|\vec p_\ell|
  (\cos\theta_P\cos\theta-\sin\theta_P\sin\theta\cos\phi)\right)W^P_7
  \nonumber\\&&\strut
  -2\left(E_\ell|\vec q\,|\cos\theta_P-q_0|\vec p_\ell|
  (\cos\theta_P\cos\theta-\sin\theta_P\sin\theta\cos\phi)\right)W^P_8
  \nonumber\\&&\strut
  -\frac1{m_b}\Big((q^2+m_\ell^2)|\vec q\,|\cos\theta_P-2q^2|\vec p_\ell|
  (\cos\theta_P\cos\theta-\sin\theta_P\sin\theta\cos\phi)\Big)W^P_9\Bigg\}.
  \qquad
\end{eqnarray}
The angle $\theta$ is reexpressed by the energy $E_\ell$ of the lepton via
\begin{equation}
\cos\theta=\frac{\vec p_\ell\cdot\vec q}{|\vec p_\ell||\vec q\,|}
  =\frac{2E_\ell q_0-q^2-m_\ell^2}{2\sqrt{E_\ell^2-m_\ell^2}\sqrt{q_0^2-q^2}}.
\end{equation}

\section{Invariant structure functions}
The Born term results to the unpolarized and polarized invariant structure
functions read
\begin{equation}
\begin{array}{lll}
W_1({\it Born\/},x^2)=\frac12(1-x^2+y^2),\quad&
W_2({\it Born\/},x^2)=2,\quad&
W_3({\it Born\/},x^2)=1,\quad\\[3mm]
W_4({\it Born\/},x^2)=0,\quad&
W_5({\it Born\/},x^2)=-1,\quad&\\[3mm]
W^P_1({\it Born\/},x^2)=-1,\quad&
W^P_2({\it Born\/},x^2)=0,\quad&
W^P_3({\it Born\/},x^2)=0,\quad\\[3mm]
W^P_4({\it Born\/},x^2)=0,\quad&
W^P_5({\it Born\/},x^2)=0,\quad&
W^P_6({\it Born\/},x^2)=1,\quad\\[3mm]
W^P_7({\it Born\/},x^2)=-1,\quad&
W^P_8({\it Born\/},x^2)=-1,\quad&
W^P_9({\it Born\/},x^2)=1
\end{array}
\end{equation}
where $y=m_c/m_b$ is the scaled mass of the $c$ quark. Note that for the Born
term results the final state $c$ quark is kinematically connected directly to
the vertex with the $b$ quark and the $W$ boson. Because of this, the on-shell
condition
\begin{equation}
\hat u=(v-x)^2-y^2=1-2x_0+x^2-y^2=0
\end{equation}
is satisfied. In this case, the scaled energy $x_0$ of the $W$ boson is
not an independent variable but is fixed by $x^2$ and $y^2$ to the value
$x_{00}=(1+x^2-y^2)/2$. Therefore, Born term results depend only on $x^2$ as
dynamical variable. This is different from radiative corrections including a
gluon as considered in the next section. In this case the hadron tensor
explicitly depends on $x_0$ (if essential, this will be indicated in the
argument) and has to be integrated over $x_0$ and over the energy of the
gluon. The invariant structure functions $W_i$ and $W^P_i$ can be written in
terms of the helicity structure functions $W_X$,
\begin{eqnarray}\label{WiGj}
W_1(x)&=&\frac12W_U,\nonumber\\[7pt]
W_2(x)&=&\frac{x^2}{2\hat x^2}(2W_L-W_U),\nonumber\\[7pt]
W_3(x)&=&-\frac1{2\hat x}W_F,\nonumber\\[7pt]
W_4(x)&=&\frac1{2x^2\hat x^2}\left(-x^2 W_U+2x_0^2W_L+2(3x_0^2-x^2)W_S
  -8x_0\hat x W_{SL}\right),\nonumber\\[7pt]
W_5(x)&=&\frac1{2\hat x^2}\Big(x_0(W_U-2W_L-2W_S)+4\hat xW_{SL}\Big),
  \nonumber\\[7pt]
W^P_1(x)&=&\frac1{2\hat x}W_{U^P},\nonumber\\[7pt]
W^P_2(x)&=&\frac1{2\hat x^3}\left(x^2(2W_{L^P}-W_{U^P})
  -4x_0\sqrt{2x^2}W_{I^P}\right),\nonumber\\[7pt]
W^P_3(x)&=&\frac1{2x^2\hat x^2}\left(-x^2W_{F^P}
  +2x_0\sqrt{2x^2}W_{A^P}+2\hat x\sqrt{2x^2}W_{SN^P}\right),\nonumber\\[7pt]
W^P_4(x)&=&\frac1{2x^2\hat x^3}
  \Big(-x^2W_{U^P}+2x_0^2W_{L^P}+2(3x_0^2-x^2)W_{S^P}
  -8x_0\hat xW_{SL^P}\nonumber\\[7pt]&&\qquad
  -4x_0\sqrt{2x^2}W_{I^P}+4\hat x\sqrt{2x^2}W_{ST^P}\Big),\nonumber\\[7pt]
W^P_5(x)&=&\frac1{2x^2\hat x^3}\Big(x^2x_0(W_{U^P}-2W_{L^P}-2W_{S^P})
  +4x^2\hat xW_{SL^P}\nonumber\\[7pt]&&\qquad
  +2(x_0^2+x^2)\sqrt{2x^2}W_{I^P}-2x_0\hat x\sqrt{2x^2}W_{ST^P}\Big),
  \nonumber\\[7pt]
W^P_6(x)&=&\frac{\sqrt{2x^2}}{\hat x}W_{I^P},\nonumber\\[7pt]
W^P_7(x)&=&-\frac{\sqrt{2x^2}}{x^2\hat x}
  \left(x_0W_{I^P}-\hat xW_{ST^P}\right),\nonumber\\[7pt]
W^P_8(x)&=&\frac{\sqrt{2x^2}}{\hat x}W_{SN^P},\nonumber\\[7pt]
W^P_9(x)&=&-\frac{\sqrt{2x^2}}{x^2\hat x}\left(\hat xW_{A^P}
  +x_0W_{SN^P}\right),
\end{eqnarray}
where $x_0=q_0/m_b$,
\begin{equation}
\hat x=\frac{|\vec q\,|}{m_b}=\sqrt{(v\cdot x)^2-x^2},
\end{equation}
and $4\hat x^2=\lambda(1,x^2,y^2)=1+x^4+y^4-2x^2-2y^2-2x^2y^2$. The helicity
structure functions can be calculated from the hadron tensor by using the
covariant helicity projection operators
\begin{eqnarray}
P_{U+L}^{\mu\nu}&=&-g^{\mu\nu}+\frac{x^\mu x^\nu}{x^2},\nonumber\\[7pt]
P_U^{\mu\nu}&=&-g^{\mu \nu}+\frac{x^\mu x^\nu}{x^2}
  -\frac{x^2}{\hat x^2}\left(v^\mu-\frac{v\cdot x}{x^2}x^\mu\right)
  \left(v^\nu-\frac{v\cdot x}{x^2}x^\nu\right),\nonumber\\[7pt]
P_L^{\mu\nu}&=&\frac{x^2}{\hat x^2}\left(v^\mu-\frac{v\cdot x}{x^2}x^\mu\right)
  \left(v^\nu-\frac{v\cdot x}{x^2}x^\nu\right),\nonumber\\[7pt]
P_F^{\mu \nu}&=&-\frac{i}{\hat x}\epsilon^{\mu\nu\rho\sigma}v_\rho x_\sigma,
  \nonumber\\[7pt]
P_I^{\mu\nu}&=&\frac{\sqrt{x^2}}{2\sqrt2\hat x}\left[
  s_T^\mu\left(v^\nu-\frac{v\cdot x}{x^2}x^\nu\right)
  +\left(v^\mu-\frac{v\cdot x}{x^2}x^\mu\right)s_T^\nu\right],\nonumber\\[7pt]
P_A^{\mu\nu}&=&-\frac{i\sqrt{x^2}}{2\sqrt2\hat x^2}\Bigg[
  \epsilon^{\mu\rho\sigma\tau}\left(v^\nu-\frac{v\cdot x}{x^2}x^\nu\right)
  -\left(v^\mu-\frac{v\cdot x}{x^2}x^\mu\right)\epsilon^{\nu\rho\sigma\tau}
  \Bigg]v_\rho x_\sigma s_{T,\tau},\nonumber\\[7pt]
P_S^{\mu\nu}&=&\frac{x^\mu x^\nu}{x^2},\nonumber\\[7pt]
P_{SL}^{\mu\nu}&=&-\frac1{2\hat x}\left(v^\mu x^\nu+x^\mu v^\nu
  -2\frac{v\cdot x}{x^2}x^\mu x^\nu\right),\nonumber\\[7pt]
P_{ST}^{\mu\nu}&=&-\frac1{2\sqrt{2x^2}}\Big(s_T^\mu x^\nu+x^\mu s_T^\nu\Big),
  \nonumber\\[7pt]
P_{SN}^{\mu\nu}&=&-\frac{i}{2\hat x\sqrt{2x^2}}
  \Big(\epsilon^{\mu\rho\sigma\tau}x^\nu-x^\mu\epsilon^{\nu\rho\sigma\tau}\Big)
  v_\rho x_\sigma s_{T,\tau}
\end{eqnarray}
and the longitudinal or transverse polarization vectors of the bottom quark,
\begin{equation}
s_L^\mu=\frac1{\hat x}\Big(x^\mu-(v\cdot x)v^\mu\Big)=(0,0,0,1),\qquad
s_T^\mu=(0,1,0,0).
\end{equation}
The helicity structure functions $W_{X^P}$ for $X=U,L,F,S,SL$ depend on the
longitudinal polarization vector while $W_{X^P}$ for $X=I,A,ST,SN$ depend on
the transverse polarization vector. Note that the relations between the
helicity structure functions $W_X$ and the invariant unpolarized and polarized
structure functions $W_i$ are expressed with coefficients depending on
$\hat x$ and $x_0$. This means that after integration over $x_0$ the relations
will mix. However, because the helicity structure functions describe angular
distributions with angles being the same for Born term and first order
contributions, in the following we will use the helicity structure functions
only. With
\begin{eqnarray}\label{Wh}
W_U&=&2W_1,\nonumber\\[7pt]
W_L&=&W_1+\frac{\hat x^2}{x^2}W_2,\nonumber\\[12pt]
W_F&=&-2\hat xW_3,\nonumber\\[7pt]
W_S&=&-W_1+\frac{x_0^2}{x^2}W_2+x^2W_4+2x_0W_5,\nonumber\\[3pt]
W_{SL}&=&\frac1{2\hat x}\left(-x_0W_1+\frac{2x_0^2-x^2}{x^2}x_0W_2
  +x_0x^2W_4+(3x_0^2-x^2)W_5\right),\nonumber\\[7pt]
W_{U^P}&=&2\hat xW^P_1,\nonumber\\[7pt]
W_{L^P}&=&\hat x\left(W^P_1+\frac{\hat x^2}{x^2}W^P_2
  +\frac{2x_0}{x^2}W^P_6\right),\nonumber\\[7pt]
W_{F^P}&=&-2\left(\hat x^2W^P_3+W^P_8+x_0W^P_9\right),\nonumber\\[7pt]
W_{I^P}&=&\frac{\hat x}{\sqrt{2x^2}}W^P_6,\nonumber\\[3pt]
W_{A^P}&=&\frac{-1}{\sqrt{2x^2}}(x_0W^P_8+x^2W^P_9),\nonumber\\[3pt]
W_{S^P}&=&\hat x\left(-W^P_1+\frac{x_0^2}{x^2}W^P_2+x^2W^P_4+2x_0W^P_5
  +\frac{2x_0}{x^2}W^P_6+2W^P_7\right),\nonumber\\[3pt]
W_{SL^P}&=&\frac12\Bigg(-x_0W^P_1+\frac{2x_0-x^2}{x^2}x_0W^P_2
  +\frac{x^2x_0}2W^P_4+(3x_0^2-x^2)W^P_5\strut\nonumber\\&&\strut\qquad\qquad
  +\frac{4x_0^2-x^2}{x^2}W^P_6+3x_0W^P_7\Bigg),\nonumber\\[3pt]
W_{ST^P}&=&\frac1{\sqrt{2x^2}}(x_0W^P_6+x^2W^P_7),\nonumber\\[3pt]
W_{SN^P}&=&\frac1{\sqrt{2x^2}}\hat xW^P_8
\end{eqnarray}
we perform the transition from the invariant structure functions to the
helicitity structure fuctions which are dealt with in the following.

\section{First order tree graph contributions}
We calculate the tree graph contributions with a finite gluon mass $m_g$
to regularize the infrared divergence which arises in the phase space
integration. To isolate the divergence it is necessary to write the
result as a sum of a singular (or soft) part $W^s({\it tree\/})$ which
contains the divergence and an IR-finite (regular) part $W^r({\it tree\/})$,
\begin{equation} \label{WGsingreg}
W({\it tree\/})=W^s({\it tree\/})+W^r({\it tree\/})
\end{equation}
($W$ stands for $W_i$ or for $W_X$). The singular part can be
written in the form
\begin{equation}
W^s({\it tree\/},x^2,x_0)=\frac{\alpha_sC_F}{4\pi}
  W({\it Born\/},x^2,x_0)S_g(x^2,x_0,\Lambda),
\end{equation}
where the scaled gluon mass $\Lambda=m_g/m_b$ is introduced. The soft gluon
factor is defined by
\begin{equation}
S_g(x^2,x_0,\Lambda)=-\Bigg[\frac1{\hat u_-}-\frac1{\hat u_+}
  -4\frac{1-x_0}{\hat u-\Lambda^2}\ln\pfrac{\hat u_+}{\hat u_-}
  +4\frac{1+x^2-2x_0-\Lambda^2}{(\hat u-\Lambda^2)^2}(\hat u_+-\hat u_-)\Bigg],
\end{equation}
where $\hat u=1+x^2-y^2-2x_0$ is the off-shell parameter and
\begin{equation}
\hat u_\pm=\frac{(1-x_0)(\hat u+\Lambda^2)\pm\sqrt{x_0^2-x^2}
  \sqrt{(\hat u-\Lambda^2)^2-4y^2\Lambda^2}}{1+x^2-2x_0}.
\end{equation}
The procedure would be much easier if the IR singular part (the logarithmic
singularity residing in $1/\hat u$) would be proportional to the proper Born
term contibution. However, this is not the case. Instead,
$W({\it Born\/},x^2,x_0)$ depends independently on $x_0$ as well because the
three-body kinematics is used. In order to get rid of this problem, we add
and subtract the two-body Born term contribution times the soft gluon factor
to obtain
\begin{equation}\label{Wplus}
W^s({\it tree\/},x^2,x_0)=\Big[W^s({\it tree\/},x^2,x_0)\Big]_+
  +\frac{\alpha_sC_F}{4\pi}W({\it Born\/},x^2)S_g(x^2,x_0,\Lambda),
\end{equation}
The term
\begin{equation}
\Big[W^s({\it tree\/},x^2,x_0)\Big]_+
  =\frac{\alpha_sC_F}{4\pi}\left(W({\it Born\/},x^2,x_0)
  -W({\it Born\/},x^2)\right)S_g(x^2,x_0,\Lambda).
\end{equation}
is IR finite, and the replacement $\Lambda\to 0$ allows for an analytical
integration of this contribution, while for the second part on the right hand
side of Eq.~(\ref{Wplus}) the Born term factor can be kept out of the
integration. This is the benefit of the plus prescription applied in
Eq.~(\ref{Wplus}). We are left with the integrated soft gluon factor defined
by
\begin{equation}\label{Ag}
A_g(x^2,\Lambda)=\int_{x_-}^{x_+}S_g(x^2,x_0,\Lambda)dx_0,
\end{equation}
where the integration limits are given by
\begin{equation}
x_-=\sqrt{x^2},\qquad x_+=\frac12\left(1+x^2-(y+\Lambda)^2\right).
\end{equation}
In the limit $\Lambda\rightarrow 0$ this leads to the analytical result
\begin{eqnarray}
A_g(x^2,\Lambda)&=&4\left\{1+2\ln\pfrac{\Lambda y}{(1-x_-)^2-y^2}-2\Li_2(x_-)
  +\Li_2\pfrac{x_-}{\eta}+\Li_2(\eta x_-)\right\}\nonumber\\[7pt]&&\strut
  +\frac2{\sqrt\lambda}\left(1-x^2+y^2\right)\Bigg\{-\ln(\omega_1)
  +\frac12\ln^2(\omega_1)-2\ln^2\pfrac{1-x_-}{1-\eta x_-}
  \nonumber\\[7pt]&&\strut\qquad
  +2\ln(\omega_1)\ln\pfrac{(\eta+1)\Lambda y}{(\eta-1)\sqrt\lambda}
  +2\Li_2(1-\omega_1)\nonumber\\[7pt]&&\strut\qquad
  -4\Li_2\pfrac{(\eta-1)x_-}{1-x_-}-4\Li_2\pfrac{(\eta-1)x_-}{\eta-x_-}\Bigg\},
\end{eqnarray}
where we have used the definitions
\begin{eqnarray}
\lambda=\lambda(1,x^2,y^2)=1+x^4+y^4-2x^2-2y^2-2x^2y^2,&&
\eta=\frac{1+x^2-y^2+\sqrt\lambda}{2x}\nonumber\\[7pt]   
\omega_1=\frac{\eta(1-\eta x_-)}{\eta-x_-}
  =\frac{1-x^2+y^2-\sqrt\lambda}{1-x^2+y^2+\sqrt\lambda},&&
x_-=\sqrt{x^2}.
\end{eqnarray}
To describe the spectrum of the charged lepton in the final state we have to
change the lower integration limit $x_-$ in the integrated soft gluon
factor~(\ref{Ag}) to $x_0(x_\ell)$ which depends on the scaled lepton
energy $x_\ell$. $A_g(x^2,\Lambda)$ then changes to the $x_\ell$-dependent
soft gluon factor
\begin{equation}
A_g(x^2,\Lambda,x_\ell)
  =\int_{x_0(x_\ell)}^{x_+}S_g(x^2,x_0,\Lambda)dx_0.
\end{equation}
The change can be performed by calculating
\begin{equation}\label{Agell}
A_g(x^2,\Lambda,x_\ell)=\int_{x_-}^{x_+}S_g(x^2,x_0,\Lambda)dx_0
  -\int_{x_-}^{x_0(x_\ell)}S_g (x^2,x_0,0)dx_0
\end{equation}
because the subtracted integral is IR finite. The limit of integration
$x_0(x_\ell)$ for a finite scaled lepton mass $\zeta=m_\ell/m_b$ is determined
by the condition
\begin{equation}
-1\leq\cos\theta=\frac{x^2x_\ell-x_0(x^2+\zeta^2)}{\sqrt{x_0^2-x^2}
  (x^2-\zeta^2)}\leq 1.
\end{equation}
This gives two solutions for $x_0$,
\begin{equation}
x_{0\pm}=\frac1{4\zeta^2}\left[x_\ell(x^2+\zeta^2)\pm\sqrt{x_\ell^2-4\zeta^2}
  \left(x^2-\zeta^2\right)\right],
\end{equation}
where the solution $x_{0-}$ is the physical one which for vanishing lepton
mass $\zeta=0$ has the limit
\begin{equation}\label{x0grenze}
x_0(x_\ell)=\frac{x^2+x_\ell^2}{2x_\ell}.
\end{equation}

The analytical result for the subtrahend in Eq.~(\ref{Agell}) is given by
\begin{eqnarray}
\lefteqn{\int_{x_-}^{x_0(x_\ell)}S_g(x^2,x_0,0)dx_0
  \ =\ \Bigg[4\ln\pfrac{(u_2-u)(uu_2-1)}{uu_2}+2\Li_2\pfrac{x_-}{u}
  +2\Li_2(ux_-)}\nonumber\\[7pt]&&
  +\frac2{\sqrt{\lambda}}\left(1-x^2+y^2\right)
  \Bigg\{\Li_2\pfrac{u}{u_2}-\Li_2\pfrac{uu_2-1}{uu_2}
  +\Li_2\pfrac{uu_2-1}{u_2(u-x_-)}\nonumber\\[7pt]&&
  -\Li_2\pfrac{u-x_-}{u_2-x_-}-\Li_2\pfrac{u_2(1-ux_-)}{u_2-x_-}
  -\Li_2\pfrac{1-u_2x_-}{1-ux_-}-\frac12\ln^2(uu_2)\nonumber\\[7pt]&&
  +\ln u\ln\pfrac{u_2-u}{u_2}+\ln(u_2)\ln(uu_2-1)
  -\ln(u-x_-)\ln\pfrac{u_2-u}{u_2-x_-}\nonumber\\[7pt]&&
  +\ln\left((u_2-u)x_-\right)\ln(1-ux_-)
  -\ln\pfrac{(uu_2-1)x_-}{u_2-x_-}\ln(1-ux_-)-\frac12\ln^2(1-ux_-)
  \nonumber\\[7pt]&&
  -\ln\pfrac{u_2}{1-u_2x_-}\ln\pfrac{uu_2-1}{1-u_2x_-}
  +\frac12\ln^2\pfrac{u_2(u-x_-)}{1-u_2x_-}-\frac12\ln^2(1-u_2x_-)
  \Bigg\}\Bigg]_{u=1}^{u(x_\ell)},\qquad
\end{eqnarray}
where $u_2=(1+x^2-y^2+\sqrt\lambda)/(2x_-)$. The upper limit depends on the
scaled lepton energy $x_\ell$ and is defined by
\begin{equation}
u(x_\ell)=\frac1{x_-}\left(x_0(x_\ell)+\sqrt{x_0(x_\ell)^2-x^2}\right),
\end{equation}
where $x_0(x_\ell)$ is the integration boundary defined in
Eq.~(\ref{x0grenze}). As the quantity $x_-$ cease from being directly related
to the lower boundary, for simplicity in the following we will skip the lower
minus sign and use the obvious notation $x=\sqrt{x^2}$ instead. Our results
presented in the main text are obtained for zero charged lepton mass,
$m_\ell=0$.

The infrared finite regular part of the tree graph contribution to the
unpolarized and polarized invariant structure functions of the decay
$b^\uparrow\rightarrow c+\ell^-+\bar\nu_\ell$ can be calculated directly
with zero gluon mass. Employing an operational notation
\begin{eqnarray}
f(x_0)\Big[S_g(x^2,x_0,\Lambda)\Big]_+&=&
  \Big(f(x_0)-f(x_{00})\Big)S_g(x^2,x_0,0)
  +f(x_{00})S_g(x^2,x_{00},\Lambda)
\end{eqnarray}
for the plus prescription where $f(x_0)$ is regular in $x_0=x_{00}$, the tree
graph contributions read explicitly
\begin{eqnarray}
\lefteqn{W_1({\it tree\/},x^2,x_0)\ =\ \frac{\alpha_sC_F}{4\pi}\Bigg\{
  \frac12(1-x^2+y^2)\left[S_g(x^2,x_0,\Lambda)\right]_+}
  \nonumber\\[7pt]&&\strut
  +\frac{1+x^2-y^2-2 x_0}{2(x_0^2-x^2)}\Bigg[9-2x^2+y^2-x_0
  +\frac12\,\frac{(1-x^2)(1-5x^2+2y^2)}{1+x^2-2x_0}
  \nonumber\\[7pt]&&\strut\qquad
  +\frac12\,\frac{(1-x^2)^3}{(1+x^2-2x_0)^2}
  +\frac{5+x^2+y^2-2x_0(2+x_0)}{\sqrt{x_0^2-x^2}}\ln(\tau)\Bigg]\Bigg\},
  \nonumber\\[12pt]
\lefteqn{W_2({\it tree\/},x^2,x_0)\ =\ \frac{\alpha_sC_F}{4\pi}\Bigg\{
  4\left[S_g(x^2,x_0,\Lambda)\right]_+}\nonumber\\[7pt]&&\strut
  +\frac1{x_0^2-x^2}\Bigg[-11-9x^2+5y^2-16x_0\nonumber\\[7pt]&&\strut\qquad
  +\frac32x\frac{(1+x)^2-y^2}{(1+x)(x_0+x)}(5+4x-x^2+y^2)
  -\frac32x\frac{(1-x)^2-y^2}{(1-x)(x_0-x)}(5-4x-x^2+y^2)
  \nonumber\\[7pt]&&\strut\qquad
  +\frac{(1-x^2)^3+2(2+x^2-3x^4)y^2+3(1+x^2)y^4}{(1-x^2)(1+x^2-2x_0)}
  -\frac{y^2(1-x^2)(1-x^2+y^2)}{(1+x^2-2x_0)^2}\Bigg]
  \nonumber\\[7pt]&&\strut
  +\frac1{4\sqrt{x_0^2-x^2}}\Bigg[32
  -\frac{(1-x^2)(5+x^2)-2(2-x^2)y^2-y^4+8(1-x^2+y^2)x_0}{x_0^2-x^2}
  \nonumber\\[7pt]&&\strut\quad
  -\frac32\,\frac{(1-x)^2-y^2}{(x_0-x)^2}(5-4x-x^2+y^2)
  -\frac32\,\frac{(1+x)^2-y^2}{(x_0+x)^2}(5+4x-x^2+y^2)\Bigg]
  \ln(\tau)\Bigg\},\nonumber\\[12pt]
\lefteqn{W_3({\it tree\/},x^2,x_0)\ =\ \frac{\alpha_sC_F}{4\pi}\Bigg\{
  -2\left[S_g(x^2,x_0,\Lambda)\right]_+}\nonumber\\[7pt]&&\strut
  +\frac1{2(x_0^2-x^2)}\Bigg[1+5x^2-y^2+14x_0
  +\frac{(1-x^2)^2y^2}{(1+x^2-2x_0)^2}\nonumber\\[7pt]&&\strut\qquad
  -\frac{(1-x^2)^2+4x^2y^2}{(1+x^2-2x_0)}
  +2x_0\frac{3-x^2+y^2-2x_0}{\sqrt{x_0^2-x^2}}
  \ln(\tau)\Bigg]\Bigg\},\nonumber\\[12pt]
\lefteqn{W_4({\it tree\/},x^2,x_0)\ =\  \frac{\alpha_sC_F}{4\pi}\Bigg\{
  \frac1{2(x_0^2-x^2)}\Bigg[-28-4x_0}\nonumber\\&&\strut\qquad
  -2\frac{(1-x^2)y^4}{(1+x^2-2x_0)^2}
  +2\frac{y^2}{1-x^2}\,\frac{6(1-x^2)+(5+x^2)y^2}{1+x^2-2x_0}
  \nonumber\\[7pt]&&\strut\qquad
  -3\frac{(1-x)^2-y^2}{x(1-x)}\frac{5-4x-x^2+y^2}{x_0-x}
  +3\frac{(1+x)^2-y^2}{x(1+x)}\frac{5+4x-x^2+y^2}{x_0+x}\Bigg]
  \nonumber\\[12pt]&&\strut
  +\frac1{4\sqrt{x_0^2-x^2}}\Bigg[
  \frac{15-36x^2+5x^4-2(6+x^2)y^2-3y^4+16x^2x_0}{x^2(x_0^2-x^2)}
  \nonumber\\[7pt]&&\strut\qquad
  -\frac32\,\frac{(1-x)^2-y^2}{x^2}\,\frac{5-4x-x^2+y^2}{(x_0-x)^2}
  -\frac32\,\frac{(1+x)^2-y^2}{x^2}\,\frac{5+4x-x^2+y^2}{(x_0+x)^2}
  \Bigg]\ln(\tau)\Bigg\},\nonumber\\[12pt]
\lefteqn{W_5({\it tree\/},x^2,x_0)\ =\  \frac{\alpha_sC_F}{4\pi}\Bigg\{
  -2\left[S_g(x^2,x_0,\Lambda)\right]_+}\nonumber\\[7pt]&&\strut
  +\frac1{2(x_0^2-x^2)}\Bigg[-7(3-x^2)-3y^2+18x_0
  +\frac{(1-x^2)y^2(1-x^2+2y^2)}{(1+x^2-2x_0)^2}
  \nonumber\\[7pt]&&\strut\qquad
  -\frac{(1-x^2)^3+2(1-x^2)(5+3x^2)y^2+4(2+x^2)y^4}{(1-x^2)(1+x^2-2x_0)}
  \nonumber\\[7pt]&&\strut\qquad
  +3\frac{(1-x)^2-y^2}{1-x}\,\frac{5-4x-x^2+y^2}{x_0-x}
  +3\frac{(1+x)^2-y^2}{1+x}\,\frac{5+4x-x^2+y^2}{x_0+x}\Bigg]
  \nonumber\\[7pt]&&\strut
  +\frac1{4\sqrt{x_0^2-x^2}}\Bigg[-8
  +\frac32\,\frac{(1-x)^2-y^2}x\,\frac{5-4x-x^2+y^2}{(x_0-x)^2}
  \nonumber\\[7pt]&&\strut\qquad
  -\frac{5-14x+7x^2+2x^3-(3+2x)y^2}{x(x_0-x)}
  -\frac32\,\frac{(1+x)^2-y^2}x\,\frac{5+4x-x^2+y^2}{(x_0+x)^2}
  \nonumber\\[7pt]&&\strut\qquad
  +\frac{5+14x+7x^2-2x^3-(3-2x)y^2}{x(x_0+x)}\Bigg]
  \ln(\tau)\Bigg\},\nonumber\\[12pt]
\lefteqn{W^P_1({\it tree\/},x^2,x_0)\ =\  \frac{\alpha_sC_F}{4\pi}\Bigg\{
  -2\left[S_g(x^2,x_0,\Lambda)\right]_+}\nonumber\\[7pt]&&\strut
  +\frac1{2(x_0^2-x^2)}\Bigg[27+7x^2-5y^2+10x_0
  -\frac{(1-x^2)^2+4x^2y^2-2y^4}{1+x^2-2x_0}\nonumber\\[7pt]&&\strut\qquad
  +\frac{(1-x^2)^2y^2}{(1+x^2-2x_0)^2}
  +\frac3x\,\frac{\left((1-x)^2-y^2\right)^2}{x_0-x}
  -\frac3x\,\frac{\left((1+x)^2-y^2\right)^2}{x_0+x}\Bigg]
  \nonumber\\[7pt]&&\strut
  +\frac1{8\sqrt{x_0^2-x^2}}\Bigg[-16
  +\frac{3(1-x)\left((1-x)^2-y^2\right)^2}{x^2(x_0-x)^2}
  +\frac{3(1+x)\left((1+x)^2-y^2\right)^2}{x^2(x_0+x)^2}
  \nonumber\\[7pt]&&\strut\qquad
  -\frac2{x^2}\,\frac{3-18x^2-x^4-6(1-x^2)y^2+3y^4
  +4x^2(3+x^2-y^2)x_0}{x_0^2-x^2}\Bigg]
  \ln(\tau)\Bigg\},\nonumber\\[12pt]
\lefteqn{W^P_2({\it tree\/},x^2,x_0)\ =\  \frac{\alpha_sC_F}{4\pi}\Bigg\{
  \frac1{2(x_0^2-x^2)}\Bigg[
  \frac{4y^2}{(1-x^2)^2}\,\frac{3(1-x^2)^2-2(3+2x^2)y^2}{1+x^2-2x_0}}
  \nonumber\\[7pt]&&\strut\qquad
  -\frac{9-6x+7x^2}{2x(1-x)^2}\,\frac{\left((1-x)^2-y^2\right)^2}{x_0-x}
  +\frac{9+6x+7x^2}{2x(1+x)^2}\,\frac{\left((1+x)^2-y^2\right)^2}{x_0+x}
  \nonumber\\[7pt]&&\strut\qquad
  +\frac{4y^4}{\left(1+x^2-2x_0\right)^2}
  -\frac{15}2\,\frac{\left((1-x)^2-y^2\right)^2}{(x_0-x)^2}
  -\frac{15}2\,\frac{\left((1+x)^2-y^2\right)^2}{(x_0+x)^2}\Bigg]
  \nonumber\\[7pt]&&\strut
  +\frac1{4\sqrt{x_0^2-x^2}}\Bigg[
  \frac{(1-x^2)(3-7x^2)-2(3+5x^2)y^2+3y^4}{4x^2(x_0^2-x^2)}
  \nonumber\\[7pt]&&\strut\qquad
  -\frac38\,\frac{(1-3x)\left((1-x)^2-y^2\right)^2}{x^2(x_0-x)^2}
  -\frac38\,\frac{(1+3x)\left((1+x)^2-y^2\right)^2}{x^2(x_0+x)^2}
  \nonumber\\[7pt]&&\strut\qquad
  -\frac{15}4\,\frac{(1-x)\left((1-x)^2-y^2\right)^2}{x(x_0-x)^3}
  +\frac{15}4\,\frac{(1+x)\left((1+x)^2-y^2\right)^2}{x(x_0+x)^3}\Bigg]
  \ln(\tau)\Bigg\},\nonumber\\[12pt]
\lefteqn{W^P_3({\it tree\/},x^2,x_0)\ =\  \frac{\alpha_sC_F}{4\pi}\Bigg\{
  \frac1{\left(x_0^2-x^2\right)}\Bigg[-1-\frac{3y^2}{1+x^2-2x_0}}
  \nonumber\\[7pt]&&\strut\qquad
  +\frac32\,\frac{1+2x}{x}\,\frac{(1-x)^2-y^2}{x_0-x}
  -\frac32\,\frac{1-2x}x\,\frac{(1+x)^2-y^2}{x_0+x}\Bigg]
  \nonumber\\[7pt]&&\strut
  +\frac1{(x_0^2-x^2)^{3/2}}\Bigg[-1-2x^2+2y^2
  +\frac34\,\frac{(1-x)(1+2x)}x\,\frac{(1-x)^2-y^2}{x_0-x}
  \nonumber\\[7pt]&&\strut\qquad
  -\frac34\,\frac{(1+x)(1-2x)}x\,\frac{(1+x)^2-y^2}{x_0+x}\Bigg]
  \ln(\tau)\Bigg\},\nonumber\\[12pt]
\lefteqn{W^P_4({\it tree\/},x^2,x_0)\ =\  \frac{\alpha_sC_F}{4\pi}\Bigg\{
  \frac{1+x^2-y^2-2x_0}{4(x_0^2-x^2)}\Bigg[
  -8\frac{(1-x^2)^2-10y^2}{(1-x^2)^2(1+x^2-2x_0)}}
  \nonumber\\[7pt]&&\strut\qquad
  -8\frac{y^2}{(1+x^2-2x_0)^2}
  -15\frac{(1-x)^2-y^2}{x^2(x_0-x)^2}-15\frac{(1+x)^2-y^2}{x^2(x_0+x)^2}
  \nonumber\\[7pt]&&\strut\qquad
  +2\frac{(1-x^2)^2(15-19x^2)-5(3-8x^2+x^4)y^2+20x^2y^2x_0}{x^2(1-x^2)^2
  (x_0^2-x^2)}\nonumber\\[7pt]&&\strut\qquad
  -6\frac{(1-x_0)(5+x^2-5y^2-10x_0+4x_0^2)}{(x_0^2-x^2)^{5/2}}\ln(\tau)
  \Bigg]\Bigg\},\nonumber\\[12pt]
\lefteqn{W^P_5({\it tree\/},x^2,x_0)\ =\  \frac{\alpha_sC_F}{4\pi}\Bigg\{
  \frac1{2\left(x_0^2-x^2\right)}\Bigg[-6
  -\frac{2y^2}{(1-x^2)^2}\,\frac{3(1-x^2)^2-4(4+x^2)y^2}{1+x^2-2x_0}}
  \nonumber\\[7pt]&&\strut\qquad
  -\frac{4y^4}{(1+x^2-2x_0)^2}
  +\frac{15}2\,\frac{\left((1-x)^2-y^2\right)^2}{x(x_0-x)^2}
  -\frac{15}2\,\frac{\left((1+x)^2-y^2\right)^2}{x(x_0+x)^2}
  \nonumber\\[7pt]&&\strut\qquad
  -\frac{(1-x)^2-y^2}{x(1-x)^2}\,\frac{4(1-x)^2(3-2x)+(3+2x)y^2}{x_0-x}
  \nonumber\\[7pt]&&\strut\qquad
  +\frac{(1+x)^2-y^2}{x(1+x)^2}\,\frac{4(1+x)^2(3+2x)+(3-2x)y^2}{x_0+x}\Bigg]
  \nonumber\\[7pt]&&\strut
  +\frac1{16\sqrt{x_0^2-x^2}}\Bigg[
  \frac{(1-x^2)(3+5x^2)-2(3-x^2)y^2+3y^4}{x^2(x_0^2-x^2)}
  \nonumber\\[7pt]&&\strut\qquad
  -\frac32\,\frac{(1-x)^2-y^2}{x^3}\,
  \frac{(1-x)(5+16x-9x^2)-(5+x)y^2}{(x_0-x)^2}\nonumber\\[7pt]&&\strut\qquad
  +\frac32\,\frac{(1+x)^2-y^2}{x^3}\,
  \frac{(1+x)(5-16x-9x^2)-(5-x)y^2}{(x_0+x)^2}\nonumber\\[7pt]&&\strut\qquad
  +\frac{15(1-x)\left((1-x)^2-y^2\right)^2}{x^2(x_0-x)^3}
  +\frac{15(1+x)\left((1+x)^2-y^2\right)^2}{x^2(x_0+x)^3}\Bigg]
  \ln(\tau)\Bigg\},\nonumber\\[12pt]
\lefteqn{W^P_6({\it tree\/},x^2,x_0)\ =\  \frac{\alpha_sC_F}{4\pi}\Bigg\{
  -2\left[S_g(x^2,x_0,\Lambda)\right]_+}\nonumber\\[7pt]&&\strut
  +\frac1{2(x_0^2-x^2)}\Bigg[13+15x^2-10y^2+16x_0
  -\frac{(1-x^2)^2+(1+3x^2)y^2+2y^4}{1+x^2-2x_0}\nonumber\\[7pt]&&\strut\qquad
  +\frac{(1-x^2)^2y^2}{(1+x^2-2x_0)^2}
  -3\frac{\left((1-x)^2-y^2\right)^2}{x_0-x}
  -3\frac{\left((1+x)^2-y^2\right)^2}{x_0+x}\Bigg]\nonumber\\[7pt]&&\strut
  +\frac1{8\sqrt{x_0^2-x^2}}\Bigg[-32
  -\frac{3(1-x)\left((1-x)^2-y^2\right)^2}{x(x_0-x)^2}
  +\frac{3(1+x)\left((1+x)^2-y^2\right)^2}{x(x_0+x)^2}
  \nonumber\\[7pt]&&\strut\qquad
  +2\frac{3+x^4-2(2+x^2)y^2+y^4+4(1-2x^2+2y^2)x_0}{x_0^2-x^2}\Bigg]
  \ln(\tau)\Bigg\},\nonumber\\[12pt]
\lefteqn{W^P_7({\it tree\/},x^2,x_0)\ =\  \frac{\alpha_sC_F}{4\pi}\Bigg\{
  2\left[S_g(x^2,x_0,\Lambda)\right]_+}\nonumber\\[7pt]&&\strut
  +\frac1{(x_0^2-x^2)}\Bigg[5(3+x^2-2y^2)-16x_0
  +\frac{(1-x^2)^2+(1+3x^2)y^2+2y^4}{1+x^2-2x_0}\nonumber\\[7pt]&&\strut\qquad
  -\frac{(1-x^2)^2y^2}{(1+x^2-2x_0)^2}
  +\frac{3\left((1-x)^2-y^2\right)^2}{x(x_0-x)}
  -\frac{3\left((1+x)^2-y^2\right)^2}{x(x_0+x)}\Bigg]
  \nonumber\\[7pt]&&\strut
  +\frac1{4\sqrt{x_0^2-x^2}}\Bigg[8
  -\frac{3-8x^2-23x^4-6(1-2x^2)y^2+3y^4+4x^2(6+x^2-y^2)x_0}{x^2(x_0^2-x^2)}
  \nonumber\\[7pt]&&\strut\qquad
  +\frac32\,\frac{(1-x)\left((1-x)^2-y^2\right)^2}{x^2(x_0-x)^2}
  +\frac32\,\frac{(1+x)\left((1+x)^2-y^2\right)^2}{x^2(x_0+x)^2}\Bigg]
  \ln(\tau)\Bigg\},\nonumber\\[12pt]
\lefteqn{W^P_8({\it tree\/},x^2,x_0)\ =\  \frac{\alpha_sC_F}{4\pi}\Bigg\{
  -2\left[S_g(x^2,x_0,\Lambda)\right]_+}\nonumber\\[7pt]&&\strut
  +\frac1{2(x_0^2-x^2)}\Bigg[3+x^2+16x_0
  -\frac{(1-x^2)^2+(3+x^2)y^2}{1+x^2-2x_0}
  +\frac{(1-x^2)^2y^2}{(1+x^2-2x_0)^2}\Bigg]\nonumber\\[7pt]&&\strut
  +\frac1{2(x_0^2-x^2)^{3/2}}\Big[1+x^2-y^2+2(3x_0-4x_0^2)\Big]
  \ln(\tau)\Bigg\},\nonumber\\[12pt]
\lefteqn{W^P_9({\it tree\/},x^2,x_0)\ =\  \frac{\alpha_sC_F}{4\pi}\Bigg\{
  2\left[S_g(x^2,x_0,\Lambda)\right]_+}\nonumber\\[7pt]&&\strut
  +\frac1{2\left(x_0^2-x^2\right)}\Bigg[-15+3x^2-4y^2-8x_0
  -\frac{(1-x^2)^2y^2}{(1+x^2-2x_0)^2}
  +\frac{(1-x^2)^2+(3+x^2)y^2}{1+x^2-2x_0}\Bigg]\nonumber\\[7pt]&&\strut
  +\frac1{4\sqrt{x_0^2-x^2}}\Bigg[8
  -\frac1x\,\frac{(1-x)(7+3x-2x^2)+(1-2x)y^2}{x_0-x}
  \nonumber\\[7pt]&&\strut\qquad
  +\frac1x\,\frac{(1+x)(7-3x-2x^2)+(1+2x)y^2}{x_0+x}\Bigg]\ln(\tau)\Bigg\},
\end{eqnarray}
where the argument $\tau$ of the logarithm is given by
\begin{equation}
\tau = \frac{1-x_0-\sqrt{x_0^2-x^2}}{1-x_0+\sqrt{x_0^2-x^2}}.
\end{equation}

\section{One loop contributions}
Finally, the contributions to one-loop corrections of unpolarized and polarized
invariant structure functions are given by
\begin{eqnarray}
\lefteqn{W_1({\it loop\/},x^2)\ =\ \frac{\alpha_sC_F}{8\pi}\Bigg\{
  -A_0(1-x^2+y^2)}\nonumber\\[7pt]&&\strut
  -\frac2{x^2}(1-y^2)(1-x^2+y^2)\ln(y)
  +\frac{\sqrt\lambda}{x^2}(1-3x^2+y^2)\ln(\omega_1)\Bigg\},\nonumber\\[7pt]
\lefteqn{W_2({\it loop\/},x^2)\ =\ \frac{\alpha_sC_F}{2\pi}\Bigg\{-A_0
  -\frac3{\sqrt\lambda}(1-x^2+y^2)\ln (\omega_1)\Bigg\},}\nonumber\\[7pt]
\lefteqn{W_3({\it loop\/},x^2)\ =\ \frac{\alpha_sC_F}{4\pi}\Bigg\{A_0
  +\frac2{x^2}(1-y^2)\ln(y)-\left[\frac1{x^2}\sqrt\lambda-\frac2{\sqrt\lambda}
  (1-x^2+y^2)\right]\ln(\omega_1)\Bigg\},}\nonumber\\[7pt]
\lefteqn{W_4({\it loop\/},x^2)\ =\ \frac{\alpha_sC_F}{2\pi x^2}\Bigg\{
  -2-\frac2{x^2}(1-2x^2-y^2)\ln(y)
  +\left[\frac{\sqrt\lambda}{x^2}-\frac1{\sqrt\lambda}(1-x^2-3y^2)\right]
  \ln (\omega_1)\Bigg\},}\nonumber\\[7pt]
\lefteqn{W_5({\it loop\/},x^2)\ =\ \frac{\alpha_sC_F}{4\pi}\Bigg\{
  A_0+\frac2{x^2}(1-y^2)+2\left(\frac\lambda{x^4}-1\right)\ln(y)}
  \nonumber\\[7pt]&&\strut
  -\frac1{x^2}(1+x^2-y^2)\left[\frac{\sqrt\lambda}{x^2}
  -\frac2{\sqrt\lambda}(1-x^2)\right]\ln(\omega_1)\Bigg\},\nonumber\\[7pt]
\lefteqn{W^P_1({\it loop\/},x^2)\ =\ \frac{\alpha_sC_F}{4\pi}\Bigg\{
  A_0+\frac2{x^2}(1-y^2)\ln(y)-\left[\frac{\sqrt\lambda}{x^2}
  -\frac2{\sqrt\lambda}(1-x^2+y^2)\right]\ln(\omega_1)\Bigg\},}\nonumber\\[7pt]
\lefteqn{W^P_2({\it loop\/},x^2)\ =\ \frac{\alpha_sC_F}{2\pi}\Bigg\{
  \frac2{x^2}\ln(y)-\frac1{\sqrt\lambda x^2}(1-x^2-y^2)\ln(\omega_1)\Bigg\},}
  \nonumber\\[7pt]
\lefteqn{W^P_3({\it loop\/},x^2)\ =\ \frac{\alpha_sC_F}{4\pi x^2}\Bigg\{
  2+\frac2{x^2}(1-2x^2-y^2)\ln(y)-\left[\frac{\sqrt\lambda}{x^2}
  -\frac{1-x^2-3y^2}{\sqrt\lambda}\right]\ln(\omega_1)\Bigg\},}\nonumber\\[12pt]
\lefteqn{W^P_4({\it loop\/},x^2)\ =\ 0,}\nonumber\\[12pt]
\lefteqn{W^P_5({\it loop\/},x^2)\ =\ \frac{\alpha_sC_F}{4\pi x^2}\Bigg\{
  2+\frac2{x^2}(1-2x^2-y^2)\ln(y)-\left[\frac{\sqrt\lambda}{x^2}
  -\frac{1}{\sqrt\lambda}(1-x^2-3y^2)\right]\ln(\omega_1)\Bigg\},}
  \nonumber\\[7pt]
\lefteqn{W^P_6({\it loop\/},x^2)\ =\ \frac{\alpha_sC_F}{8\pi}\Bigg\{
  2A_0+\frac2{x^2}(1+x^2-y^2)\ln(y)-\left[\frac{\sqrt\lambda}{x^2}
  -\frac4{\sqrt\lambda}(1-x^2+y^2)\right]\ln(\omega_1)\Bigg\},}\nonumber\\[7pt]
\lefteqn{W^P_7({\it loop\/},x^2)\ =\ \frac{\alpha_sC_F}{8\pi}\Bigg\{
  -2A_0-\frac2{x^2}(1-x^2-y^2)-\frac2{x^2}
  \left(\frac\lambda{x^2}+1+x^2-3y^2\right)\ln(y)}\nonumber\\[7pt]&&\strut
  +\frac1{x^2}\left[\frac{\sqrt\lambda}{x^2}(1+2x^2-y^2)
  -\frac2{\sqrt\lambda}(1-x^2+y^2)(1+x^2-y^2)\right]\ln(\omega_1)\Bigg\},
  \nonumber\\[7pt]
\lefteqn{W^P_8({\it loop\/},x^2)\ =\ \frac{\alpha_sC_F}{8\pi}\Bigg\{
  2A_0-4-\frac2{x^2}(1-3x^2-y^2)\ln(y)+\left[\frac{\sqrt\lambda}{x^2}
  +\frac4{\sqrt\lambda}(1-x^2+y^2)\right]\ln (\omega_1)\Bigg\},}\nonumber\\[7pt]
\lefteqn{W^P_9({\it loop\/},x^2)\ =\ \frac{\alpha_sC_F}{8\pi}\Bigg\{
  -2A_0+\frac2{x^2}(1+x^2-y^2)+\frac2{x^2}\left[\frac\lambda{x^2}
  +1-3x^2+5y^2\right]\ln(y)}\nonumber\\[7pt]&&\strut
  -\frac1{\sqrt\lambda x^4}\Big[(1-x^2)(1-x^2+4x^4)-(1+x^2)(3-7x^2)y^2
  +(3-2x^2)y^4-y^6\Big]\ln(\omega_1)\Bigg\}.\nonumber\\
\end{eqnarray}
The IR divergent part is proportional to the product of the Born result
times the factor
\begin{eqnarray}
A_0(x^2,\Lambda)&=&\frac2{\sqrt\lambda}(1-x^2+y^2)
  \Bigg\{-2\Li_2(1-\omega_2)+2\Li_2(1-\omega_3)
  -\ln(\omega_1)\ln\pfrac{y}{\Lambda^2}\nonumber\\[7pt]&&\strut
  -\ln\left(\frac12\left(1-x^2+y^2+\sqrt\lambda\right)\right)
  \ln(\omega_2\omega_3)
  \Bigg\}-4\left[\ln\pfrac{y}{\Lambda^2}-2\right],
\end{eqnarray}
where
\begin{equation}
\omega_2=\frac{1+x^2-y^2-\sqrt\lambda}{1+x^2-y^2+\sqrt\lambda},\qquad
\omega_3=\frac{1-x^2-y^2-\sqrt\lambda}{1-x^2-y^2+\sqrt\lambda}.
\end{equation}

\section{Integrated helicity structure functions}
In this section we integrate the invariant structure functions
$W_X({\it tree\/},x^2,x_0)$ over the scaled energy $x_0$ of the $W$ boson and
combine it with the Born term and loop results,
\begin{eqnarray}
W_X({\it incl\/},x^2)&=&W_X({\it Born\/},x^2)
  \left(1+\frac{\alpha_sC_F}{4\pi}A(x^2)\right)\nonumber\\&&\strut
    +W_X^f({\it loop\/},x^2)+\int W_X^f({\it tree\/},x^2,x_0)dx_0
\end{eqnarray}
($X\in\{U,L,F,S,SL,U^P,L^P,F^P,I^P,A^P,S^P,SL^P,ST^P,SN^P\}$), where
$W_X^f({\it loop\/},x^2)$ and $W_X^f({\it tree\/},x^2,x_0)$ contain only
finite parts where the contributions of $A_g(x^2,\Lambda)$ and
$A_0(x^2,\Lambda)$ are skipped. In calculating the difference $A_g-A_0$,
the IR sigularities cancel according to the Kinoshita--Lee--Nauenberg theorem,
and the IR-finite factor $A$ is defined by
\begin{eqnarray} \label{afaktor}
A (x^2) &=& \lim_{\Lambda\rightarrow 0} \Big[ A_g (x^2, \Lambda)
     -A_0 (x^2, \Lambda) \Big] \nonumber \\ [1.5mm]
     &=& 4 \left\{ \ln (y)-1+2\ln
     \left[ \frac{y}{(1-x)^2-y^2}\right]
     -2\Li_2 (x)+\Li_2 \left(\frac{x}{\eta}\right)
     +\Li_2 (\eta x)\right\} \\ [1.5mm]
     &+& \frac{2}{\sqrt\lambda} \left(1-x^2+y^2\right)
     \Bigg\{-\, \ln (\omega_1)+\frac12\ln^2 (\omega_1)
     -2\ln^2 \left(\frac{1-x}{1-\eta x}\right)
     \nonumber \\ [1.5mm]
     &&+\ln (\omega_1) \left[ 2\ln \left(\frac{\eta+1}{\eta-1}
     y\right)+\ln \left(\frac{y}{\lambda}\right)\right]
     +\ln \left[ \frac12 \left(1-x^2+y^2
     +\sqrt\lambda\right)\right] \ln (\omega_2\omega_3) \nonumber \\ [1.5mm]
     &&+4 \ln (\eta)\ln \left(\frac{\eta-x}{\eta}\right)
     +2\Li_2 (\eta x)-2\Li_2
     \left(\frac{x}{\eta}\right) \nonumber \\ [1.5mm]
     &&-4\Li_2 \left[ \frac{(\eta-1)x}{1-x}\right]
     -4\Li_2 \left[ \frac{(\eta-1)x}{\eta-x}\right]
     \Bigg\},
\end{eqnarray}
where the dilogarithmic identity
\begin{eqnarray}
\Li_2 (1-\omega_1)+\Li_2 (1-\omega_2)-\Li_2 (1-\omega_3)
     &=& \Li_2 (\eta x)-\Li_2 \left(\frac{x}{\eta}\right)
     +2\ln (\eta)\ln \left(\frac{\eta-x}{\eta}\right)
     \nonumber \\ [1mm]
\end{eqnarray}
is employed. According to Eqs.~(\ref{Wh}), the unintegrated inclusive helicity
structure functions $W_X({\it tree\/},x^2,x_0)$ are linear combinations
of the invariant structure functions $W_i({\it tree\/},x^2,x_0)$. In the
following we discuss the results for the differential decay rates with respect
to $x^2$ and three angles $\theta_P$, $\theta$ and $\phi$ which are defined in
Figure~\ref{kinematics}. The helicity structure functions are the coefficients
of the angular dependence in the angular decay distribution of the decay
process.

\begin{figure}[t]\begin{center}
\epsfig{figure=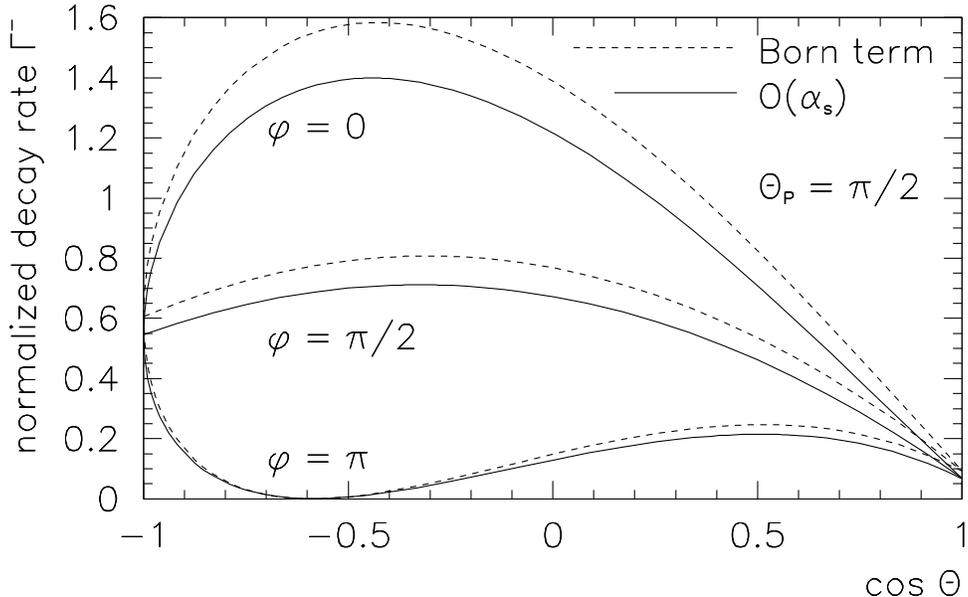, scale=0.8}
\caption{\label{figphi}Dependence of the normalized decay rate $\hat\Gamma^-$
on the cosine of the polar angle $\theta$ for azimuthal angles $\phi=0$,
$\phi=\pi/2$ and $\phi=\pi$. The polarization of the $b$ quark is orthogonal to
the momentum of the $W$ boson, $\theta_P=\pi/2$.}
\end{center}\end{figure}

The angular decay distributions of the process
$b^\uparrow\rightarrow c+\ell^-+\bar\nu_\ell$ into leptons with negative
or positive helicity, respectively, are given by
\begin{eqnarray}
\frac{d\Gamma^-}{dx^2d\cos\theta\,d\cos\theta_Pd\phi}
     &=& \frac{\Gamma_b}{4 \pi} \\ [3mm]
     && \hspace{-4.5cm} \times \left[ \frac{3}{8} \left(
     \frac{d\hat{\Gamma}_U^-}{dx^2}
     +\frac{d\hat{\Gamma}_{U^P}^-}{dx^2}P\cos\theta_P\right) \left(
     1+\cos^2 \theta\right)+\frac{3}{4} \left(
     \frac{d\hat{\Gamma}_L^-}{dx^2}
     +\frac{d\hat{\Gamma}_{L^P}^-}{dx^2}P\cos\theta_P\right) \sin^2 \theta
    \right.\nonumber\\[7pt]
     && \hspace{-3.6cm}+\left. \frac{3}{4} \left(
     \frac{d\hat{\Gamma}_F^-}{dx^2}
     +\frac{d\hat{\Gamma}_{F^P}^-}{dx^2}P\cos\theta_P\right) \cos\theta\right.
    \nonumber\\[7pt]
     && \hspace{-3.6cm}+\left. \frac{3}{\sqrt{2}}
     \frac{d\hat{\Gamma}_{I^P}^-}{dx^2}P\sin\theta_P
     \sin\theta\cos\theta\cos\phi
     +\frac{3}{\sqrt{2}}\,\frac{d\hat{\Gamma}_{A^P}^-}{dx^2}
     P\sin\theta_P\sin\theta\cos\phi\right] \nonumber \\ [5mm]
\frac{d\Gamma^+}{dx^2d\cos\theta\,d\cos\theta_Pd\phi}
     &=& \frac{\Gamma_b}{4 \pi}\nonumber\\[7pt]
     && \hspace{-4.5cm} \times \left[ \frac{3}{4} \left(
     \frac{d\hat{\Gamma}_U^+}{dx^2}
     +\frac{d\hat{\Gamma}_{U^P}^+}{dx^2}P\cos\theta_P\right) \sin^2 \theta
     +\frac32 \left(\frac{d\hat{\Gamma}_L^+}{dx^2}
     +\frac{d\hat{\Gamma}_{L^P}^+}{dx^2}P\cos\theta_P
    \right) \cos^2 \theta\right.\nonumber\\[7pt]
     && \hspace{-3.6cm}+\, \left. \frac32 \left(
     \frac{d\hat{\Gamma}_S^+}{dx^2}
     +\frac{d\hat{\Gamma}_{S^P}^+}{dx^2}P\cos\theta_P\right)+3 \left(
     \frac{d\hat{\Gamma}_{SL}^+}{dx^2}
     +\frac{d\hat{\Gamma}_{SL^P}^+}{dx^2}P\cos\theta_P\right) \cos\theta
    \right.\nonumber\\[7pt]
     && \hspace{-3.6cm}+\left. 3\sqrt{2}
     \frac{d\hat{\Gamma}_{ST^P}^+}{dx^2}P\sin\theta_P
     \sin\theta\cos\phi+3\sqrt{2}
     \frac{d\hat{\Gamma}_{I^P}^+}{dx^2}P\sin\theta_P
     \sin\theta\cos\theta\cos\phi\right] \nonumber
\end{eqnarray}
where
\begin{equation} \label{Gammab}
\Gamma_b = \frac{G_F^2m_b^5|V_{b c}|^2}{192 \pi^3}
\end{equation}
is the total decay rate of the bottom quark in the limit $m_c\rightarrow 0$.
The angular decay distributions for the transverse components of the spin for
the charged lepton read
\begin{eqnarray}
\frac{d\Gamma^x}{dx^2d\cos\theta\,d\cos\theta_Pd\phi}
     &=& \frac{\Gamma_b}{4 \pi} \\ [3mm]
     && \hspace{-5cm} \times \left\{ \frac{3}{2\sqrt{2}} \left(
     \frac{d\hat{\Gamma}_F^x}{dx^2}+\frac{d\hat{\Gamma}_{F^P}^x}{dx^2}
     P\cos\theta_P\right) \sin\theta\right.\nonumber\\[7pt]
     && \hspace{-4.1cm} \left.-\frac{3}{2\sqrt{2}} \left[
     \frac{d\hat{\Gamma}_U^x}{dx^2}+\frac{d\hat{\Gamma}_{U^P}^x}{dx^2}
     P\cos\theta_P-2 \left(
     \frac{d\hat{\Gamma}_L^x}{dx^2}+\frac{d\hat{\Gamma}_{L^P}^x}{dx^2}
     P\cos\theta_P\right)\right]
     \sin\theta\cos\theta\right.\nonumber\\[7pt]
     && \hspace{-4.1cm} \left.+\frac{3}{\sqrt{2}} \left(
     \frac{d\hat{\Gamma}_{SL}^x}{dx^2}
     +\frac{d\hat{\Gamma}_{SL^P}^x}{dx^2}P\cos\theta_P\right)
     \sin\theta-3\frac{d\hat{\Gamma}_{I^P}^x}{dx^2}
     P\sin\theta_P\cos (2\theta)\cos\phi
    \right.\nonumber\\[7pt]
     && \hspace{-4.1cm} \left.-3
     \left(\frac{d\hat{\Gamma}_{A^P}^x}{dx^2}
     +\frac{d\hat{\Gamma}_{ST^P}^x}{dx^2}\right)
     P\sin\theta_P\cos\theta\cos\phi
     -3\frac{d\hat{\Gamma}_{SN^P}^x}{dx^2}
     P\sin\theta_P\cos\phi\right\} \nonumber \\ [5mm]
\frac{d\Gamma^y}{dx^2d\cos\theta\,d\cos\theta_P
d\phi} &=& \frac{\Gamma_b}{4 \pi}\nonumber\\[7pt]
     && \hspace{-5cm} \times \left[ 3 \left(
     \frac{d\hat{\Gamma}_{A^P}^y}{dx^2}
     +\frac{d\hat{\Gamma}_{ST^P}^y}{dx^2}\right)
     P\sin\theta_P \sin\phi+3
     \left(\frac{d\hat{\Gamma}_{I^P}^y}{dx^2}
     +\frac{d\hat{\Gamma}_{SN^P}^y}{dx^2}\right)
     P\sin\theta_P\cos\theta\sin\phi\right] \nonumber.
\end{eqnarray}
The reduced rates are given by
\begin{eqnarray} \label{gamma}
\frac{d\hat{\Gamma}_X^-}{dx^2}=2\frac{(x^2-\zeta^2)^2}{x^2}T_X(x^2),&&
\frac{d\hat{\Gamma}_X^x}{dx^2}=\frac{2\zeta}{\sqrt{2x^2}}
  \frac{(x^2-\zeta^2)^2}{x^2}T_X(x^2),\nonumber\\
\frac{d\hat{\Gamma}_X^+}{dx^2}=\frac{\zeta^2}{x^2}\,
  \frac{(x^2-\zeta^2)^2}{x^2}T_X(x^2),&&
\frac{d\hat{\Gamma}_X^y}{dx^2}=\frac{2\zeta}{\sqrt{2x^2}}
  \frac{(x^2-\zeta^2)^2}{x^2}T_X(x^2)
\end{eqnarray}
with
\begin{eqnarray}
T_U &=& 2 \left(1-x^2+y^2\right) \sqrt\lambda
     -\frac{\alpha_sC_F}{4\pi} \times 2
     \bigg\{ 4 \left(1-x^2+y^2\right)^2 {\cal{N}}_1 \nonumber \\ [1.5mm]
     &&+\frac{1}{x}\Big[ (1-x)^2-y^2 \Big]
     \Big[ (1-x)(5+x)+y^2 \Big]{\cal{N}}_2 \nonumber \\ [1.5mm]
     &&-\frac{1}{x}\Big[ (1+x)^2-y^2 \Big]\Big[
     (1+x)(5-x)+y^2 \Big]{\cal{N}}_3 \nonumber \\ [1.5mm]
     &&+\frac{2}{x^2}\sqrt\lambda \left(1-x^2+y^2\right)
     \left(1-6x^2-y^2\right) \ln (y)
     -8\sqrt\lambda \left(1-x^2+y^2\right) \ln \left(
     \frac{x}{\lambda}\right) \nonumber \\ [1.5mm]
     &&-\frac{1}{x^2}\Big[ \left(1-x^2\right)^2
     \left(1-6x^2\right) \nonumber \\ [1.5mm]
     && \hspace{1cm}-\left(1+4x^2-3x^4\right) y^2-\left(
     1+2x^2\right) y^4+y^6 \Big]\ln (\omega_1)
     \nonumber \\ [1.5mm]
     &&-4\Big[ 7+3x^2
     -\left(4-5x^2\right) y^2-3y^4 \Big]\ln (\eta)
     +\sqrt\lambda \left(19+x^2-5y^2\right) \! \bigg\}
  \\[12pt]
T_{U^P} &=&-\, 2\lambda+\frac{\alpha_sC_F}{4\pi} \times 2
     \bigg\{ 4\sqrt\lambda \left(1-x^2+y^2\right)
     {\cal{N}}_4 \nonumber \\ [1.5mm]
     &&-4\Big[ 11+3x^2+x^4-2 \left(3+x^2\right) y^2
     +y^4 \Big]{\cal{N}}_5 \nonumber \\ [1.5mm]
     &&+\frac{2}{x^2}\lambda \left(1-6x^2-y^2\right)
     \ln (y)
     +8\lambda\ln \left[ (1-x)^2-y^2\right]
     \nonumber \\ [1.5mm]
     &&-\frac{\sqrt\lambda}{x^2}\Big[ 7+21x^2+2x^4
     -\left(8+3x^2\right) y^2+y^4 \Big] \ln (\omega_1)
     \nonumber \\ [1.5mm]
     &&-\frac{4}{x^2}\Big[ \left(1-x^2\right)
     \left(3+14x^2-2x^4\right) \nonumber \\ [1.5mm]
     && \hspace{1cm}-\left(6-7x^2-x^4
    \right) y^2+\left(3-x^2\right) y^4 \Big]\ln \left(
     \frac{1-x}{y}\right) \nonumber \\ [1.5mm]
     &&-\frac{1}{x}\Big[ (1-x)^2-y^2 \Big]\Big[
     12-55x+6x^2-x^3-3 \left(4+x\right) y^2 \Big] \bigg\}
  \\[12pt]
T_L &=& \frac{1}{x^2}\Big[ \lambda+x^2 \left(1-x^2+y^2
    \right) \Big]\sqrt\lambda-\frac{\alpha_sC_F}{4\pi}
     \nonumber \\ [1.5mm]
     && \times \frac{1}{x^2}\bigg\{ 4 \left(1-x^2+y^2\right)
     \Big[ \lambda+x^2 \left(1-x^2+y^2\right) \Big]
     {\cal{N}}_1 \nonumber \\ [1.5mm]
     &&-2x\Big[ (1-x)^2-y^2 \Big]\Big[ (1-x)(5+x)
     +y^2 \Big]{\cal{N}}_2 \nonumber \\ [1.5mm]
     &&+2x \Big[ (1+x)^2-y^2 \Big]\Big[ (1+x)(5-x)
     +y^2 \Big]{\cal{N}}_3 \nonumber \\ [1.5mm]
     &&-2\sqrt\lambda\Big[ 5 \left(1-x^2\right)
     -\left(12+7x^2\right) y^2+7y^4 \Big]
     \ln (y) \nonumber \\ [1.5mm]
     &&-8\sqrt\lambda
     \Big[ 1-x^2-\left(2+x^2\right) y^2+y^4 \Big]
     \ln \left(\frac{x}{\lambda}\right) \nonumber \\ [1.5mm]
     &&+\Big[ 5 \left(1-x^2\right)^2-\left(3+20x^2-x^4
    \right) y^2+\left(9-2x^2\right) y^4+y^6 \Big]\ln (\omega_1)
     \nonumber \\ [1.5mm]
     &&-8 \left(1+x^2-y^2\right) \Big[ 1-7x^2-\left(
     2+x^2\right) y^2+y^4 \Big]\ln (\eta) \nonumber \\ [1.5mm]
     &&-\sqrt\lambda\Big[ 5+47x^2-4x^4-\left(
     22+x^2\right) y^2+5y^4 \Big] \bigg\}
  \\[12pt]
T_{L^P} &=& \frac{1}{x^2} \left(1-y^2\right) \lambda
     -\frac{\alpha_sC_F}{4\pi} \times \frac{1}{x^2}
     \bigg\{ 4\sqrt\lambda \left(1-y^2\right)
     \left(1-x^2+y^2\right) {\cal{N}}_4 \nonumber \\ [1.5mm]
     &&-4\Big[ 2+22x^2+11x^4-\left(5+12x^2+x^4
    \right) y^2+2 \left(2+x^2\right) y^4-y^6 \Big]{\cal{N}}_5
     \nonumber \\ [1.5mm]
     &&-2\lambda \left(5-7y^2\right) \ln (y)
     +8\lambda \left(1-y^2\right) \ln
     \left[ (1-x)^2-y^2\right] \nonumber \\ [1.5mm]
     &&-\sqrt\lambda\Big[ 17+53x^2-\left(18+x^2
    \right) y^2+y^4 \Big]\ln (\omega_1) \nonumber \\ [1.5mm]
     &&-4\Big[ \left(1-x^2\right) \left(11+24x^2\right)
     -\left(13-15x^2\right) y^2+2y^4 \Big]\ln \left(
     \frac{1-x}{y}\right) \nonumber \\ [1.5mm]
     &&+\Big[ (1-x)^2-y^2 \Big] \nonumber \\ [1.5mm]
     && \hspace{1cm}\times \Big[ 15-22x+105x^2
     -24x^3+4x^4-\left(12-22x+x^2\right) y^2
     -3y^4 \Big] \bigg\}
  \\[12pt]
T_S &=& \frac{1}{x^2} \Big[ \lambda+x^2 \left(1-x^2+y^2
    \right) \Big]\sqrt\lambda-\frac{\alpha_sC_F}{4\pi}
     \nonumber \\ [1.5mm]
     && \times \frac{1}{x^2}\bigg\{ 4 \left(1-x^2+y^2\right)
     \Big[ \lambda+x^2 \left(1-x^2+y^2\right) \Big]
     {\cal{N}}_1 \nonumber \\ [1.5mm]
     &&-\frac{2}{x^2}\sqrt\lambda
     \Big[ \left(1-x^2\right) \left(2+3x^2
    \right)-3 \left(2+4x^2+3x^4\right) y^2
     \nonumber \\ [1.5mm]
     && \hspace{1cm}+\left(6+11x^2\right) y^4
     -2y^6 \Big]\ln (y)
     \nonumber \\ [1.5mm]
     &&-8\sqrt\lambda
     \Big[ 1-x^2-\left(2+x^2\right) y^2+y^4 \Big]
     \ln \left(\frac{x}{\lambda}\right)
     \nonumber \\ [1.5mm]
     &&+\frac{1}{x^2}\Big[ \left(1-x^2\right)^2
     \left(2+3x^2\right)-\left(8-3x^2+4x^4-3x^6
    \right) y^2 \nonumber \\ [1.5mm]
     && \hspace{1cm}+3 \left(4+5x^2\right) y^4
     -\left(8+5x^2\right) y^6+2y^8 \Big]\ln (\omega_1)
     \nonumber \\ [1.5mm]
     &&-8 \left(1-y^2\right) \Big[ 1-x^2-\left(2+x^2
    \right) y^2+y^4 \Big]\ln (\eta) \nonumber \\ [1.5mm]
     &&-3\sqrt\lambda\Big[ 3 \left(1-x^2\right)
     -\left(10+3x^2\right) y^2+3y^4 \Big] \bigg\}
  \\[12pt]
T_{S^P} &=& \frac{1}{x^2} \left(1-y^2\right) \lambda
    -\frac{\alpha_sC_F}{4\pi} \times \frac{1}{x^2} \left(1-y^2\right)
     \bigg\{ 4\sqrt\lambda \left(1-x^2+y^2\right)
     {\cal{N}}_4 \nonumber \\ [1.5mm]
     &&-4\Big[ 2+x^4-\left(3 +2x^2\right) y^2+y^4
     \Big]{\cal{N}}_5 \nonumber \\ [1.5mm]
     &&-\frac{2}{x^2}\,\frac{1}{1-y^2}\lambda
     \Big[ 2+3x^2-\left(4+9x^2\right) y^2+2y^4 \Big]
     \ln (y)
     +8\lambda\ln \left[ (1-x)^2-y^2\right]
     \nonumber \\ [1.5mm]
     &&+\frac{\sqrt\lambda}{x^2}\Big[ 2-9x^2+x^4
     -\left(4+3x^2\right) y^2+2y^4 \Big]\ln (\omega_1)
     \nonumber \\ [1.5mm]
     &&-4\Big[ \left(1-x^2\right) \left(5-2x^2\right)
     +2 \left(2-x^2\right) y^2 \Big]\ln \left(\frac{1-x}{y}
    \right) \nonumber \\ [1.5mm]
     &&+\Big[ (1-x)^2-y^2 \Big] \left(11-6x-7x^2
     +7y^2\right) \! \bigg\}
  \\[12pt]
T_F &=&-\, 2\lambda+\frac{\alpha_sC_F}{4\pi} \times 2\bigg\{
     4\sqrt\lambda \left(1-x^2+y^2\right)
     {\cal{N}}_4 \nonumber \\ [1.5mm]
     &&+4 \left(1+3x^2-x^4+2x^2y^2-y^4\right)
     {\cal{N}}_5 \nonumber \\ [1.5mm]
     &&+\frac{2}{x^2}\lambda \left(1-6x^2-y^2\right) \ln (y)
     +8\lambda\ln \left[ (1-x)^2-y^2\right]
     \nonumber \\ [1.5mm]
     &&-\frac{\sqrt\lambda}{x^2}\Big[ 1-9x^2+2x^4
     -\left(2+3x^2\right) y^2+y^4 \Big]\ln (\omega_1)
     \nonumber \\ [1.5mm]
     &&+4\Big[ \left(1-x^2\right) \left(1+2x^2\right)
     -\left(1+x^2\right) y^2 \Big]\ln \left(\frac{1-x}{y}\right)
     \nonumber \\ [1.5mm]
     &&-2 \left[ (1-x)^2-y^2\right]
     \left(3-4x-3y^2\right) \! \bigg\}
  \\[12pt]
T_{F^P} &=& 2\sqrt\lambda \left(1-x^2+y^2\right)
     -\frac{\alpha_sC_F}{4\pi} \times 2
     \bigg\{ 4 \left(1-x^2+y^2\right)^2 {\cal{N}}_1 \nonumber \\ [1.5mm]
     &&+\frac{2}{x}(1-x)(1+2x) \left[ (1-x)^2-y^2
    \right] {\cal{N}}_2 \nonumber \\ [1.5mm]
     &&-\frac{2}{x}(1+x)(1-2x)\Big[
     (1+x)^2-y^2 \Big]{\cal{N}}_3 \nonumber \\ [1.5mm]
     &&+\frac{2}{x^2}\sqrt\lambda \left(1-x^2+y^2\right)
     \left(1-6x^2-y^2\right) \ln (y)
     -8\sqrt\lambda \left(1-x^2+y^2\right) \ln \left(
     \frac{x}{\lambda}\right) \nonumber \\ [1.5mm]
     &&-\frac{1}{x^2}\Big[ \left(1-x^2\right)^2
     \left(1-6x^2\right) \nonumber \\ [1.5mm]
     && \hspace{1cm}-\left(1-8x^2-3x^4\right) y^2
     -\left(1+4x^2\right) y^4+y^6 \Big]\ln (\omega_1)
     \nonumber \\ [1.5mm]
     &&-4\Big[ 4-9x^2-\left(2-5x^2\right) y^2
     -2y^4 \Big]\ln (\eta)
     -2\sqrt\lambda \left(4+x^2+2y^2\right) \! \bigg\}
  \\[12pt]
T_{I^P} &=&-\, \frac{1}{\sqrt{2}x}\lambda
     +\frac{\alpha_sC_F}{4\pi} \times \frac{1}{\sqrt{2}x}\bigg\{
     4\sqrt\lambda \left(1-x^2+y^2\right)
     {\cal{N}}_4 \nonumber \\ [1.5mm]
     &&-2\Big[ 7+15x^2+4x^4-\left(11+8x^2
    \right) y^2+4y^4 \Big]{\cal{N}}_5 \nonumber \\ [1.5mm]
     &&+\frac{1}{x^2}\lambda \left(1-11x^2-y^2\right)
     \ln (y)
     +8\lambda\ln \left[ (1-x)^2-y^2\right]
     \nonumber \\ [1.5mm]
     &&-\frac{\sqrt\lambda}{2x^2}\Big[ 1+30x^2+21x^4
     -2 \left(1+11x^2\right) y^2+y^4 \Big]\ln (\omega_1)
     \nonumber \\ [1.5mm]
     &&-2\Big[ \left(1-x^2\right) \left(21+5x^2\right)
     -\left(11-15x^2\right) y^2-4y^4 \Big]\ln \left(
     \frac{1-x}{y}\right) \nonumber \\ [1.5mm]
     &&+2\Big[ (1-x)^2-y^2 \Big]
     \left(12-7x+12x^2-9y^2\right) \! \bigg\}
  \\[12pt]
T_{A^P} &=& \frac{1}{\sqrt{2}x}\sqrt\lambda \left(1-x^2-y^2
    \right)-\frac{\alpha_sC_F}{4\pi} \times \frac{1}{\sqrt{2}x}
     \bigg\{ 4 \left(1-x^2+y^2\right) \left(1-x^2-y^2\right)
     {\cal{N}}_1 \nonumber \\ [1.5mm]
     &&-(1-x)(1+2x) \left[
     (1-x)^2-y^2\right] {\cal{N}}_2 \nonumber \\ [1.5mm]
     &&-(1+x)(1-2x) \left[ (1+x)^2-y^2\right]
     {\cal{N}}_3 \nonumber \\ [1.5mm]
     &&+\frac{1}{x^2}\sqrt\lambda
     \Big[ \left(1-x^2\right) \left(1-11x^2\right)
     -2 \left(1-8x^2\right) y^2+y^4 \Big]\ln (y)
     \nonumber \\ [1.5mm]
     &&-8\sqrt\lambda \left(1-x^2-y^2\right)\ln \left(
     \frac{x}{\lambda}\right) \nonumber \\ [1.5mm]
     &&-\frac{1}{2x^2}\Big[ \left(1-x^2\right)^2
     \left(1-11x^2\right) \nonumber \\ [1.5mm]
     && \hspace{1cm}-\left(1+x^2\right) \left(3-11x^2
    \right) y^2+\left(3-7x^2\right) y^4-y^6 \Big]\ln (\omega_1)
     \nonumber \\ [1.5mm]
     &&-2\Big[ \left(1+x^2\right) \left(4-7x^2\right)
     -\left(8-7x^2\right) y^2+4y^4 \Big]\ln (\eta)
     \nonumber \\ [1.5mm]
     &&-2\sqrt\lambda \left(1+2x^2-4y^2\right) \!
     \bigg\}.
  \\[12pt]
T_{SL} &=& \frac{1}{x^2} \left(1-y^2\right) \lambda
     -\frac{\alpha_sC_F}{4\pi} \times \frac{1}{x^2}
     \bigg\{ 4\sqrt\lambda \left(1-y^2\right)
     \left(1-x^2+y^2\right)
     {\cal{N}}_4 \nonumber \\ [1.5mm]
     &&+4\Big[ 1+5x^2-x^4-\left(1+x^2-x^4\right) y^2
     -\left(1+2x^2\right) y^4+y^6 \Big]{\cal{N}}_5
     \nonumber \\ [1.5mm]
     &&-\frac{2}{x^2}\lambda\Big[ 1+4x^2
     -2 \left(1+4x^2\right) y^2+y^4 \Big]\ln (y)
     \nonumber \\ [1.5mm]
     &&+\frac{\sqrt\lambda}{x^2}\Big[ 1+7x^2+2x^4
     -\left(3+8x^2\right) y^2+\left(3+x^2\right) y^4
     -y^6 \Big] \ln (\omega_1) \nonumber \\ [1.5mm]
     &&+8 \left(1-y^2\right) \lambda
     \ln \left[ (1-x)^2-y^2\right] \nonumber \\ [1.5mm]
     &&+4\Big[ \left(1-x^2\right) \left(2+3x^2\right)
     -\left(7-4x^2-x^4\right) y^2-\left(1+x^2\right) y^4
     \Big] \ln \left(\frac{1-x}{y}\right) \nonumber \\ [1.5mm]
     &&-\left[ (1-x)^2-y^2\right] \Big[
     7-10x+13x^2-\left(26+2x+7x^2\right) y^2
     +7y^4 \Big] \bigg\}
  \\[12pt]
T_{SL^P} &=& \frac{\sqrt\lambda}{x^2}
     \Big[ \lambda+x^2 \left(1-x^2+y^2\right) \! \Big]
     -\frac{\alpha_sC_F}{4\pi} \nonumber \\ [1.5mm]
     && \times \frac{1}{x^2}\bigg\{ 4 \left(1-x^2+y^2\right)
     \Big[ \lambda+x^2 \left(1-x^2+y^2\right) \Big]{\cal{N}}_1
     \nonumber \\ [1.5mm]
     &&-2 \left[ (1-x)^2-y^2\right] \left[ 3(1-x)
     -(3-2x)y^2\right] {\cal{N}}_2 \nonumber \\ [1.5mm]
     &&-2 \left[ (1+x)^2-y^2\right] \left[ 3(1+x)
     -(3+2x)y^2\right] {\cal{N}}_3 \nonumber \\ [1.5mm]
     &&-\frac{2}{x^2}\sqrt\lambda\Big[ \left(1-x^2\right)
     \left(1+4x^2\right)-\left(3+12x^2+8x^4\right) y^2
     \nonumber \\
     && \hspace{1.9cm}+3 \left(1+3x^2\right) y^4-y^6 \Big]
     \ln (y) \nonumber \\ [1.5mm]
     &&-8\sqrt\lambda\Big[ 1-x^2-\left(2+x^2\right) y^2
     +y^4 \Big]\ln \left(\frac{x}{\lambda}\right)
     \nonumber \\ [1.5mm]
     &&+\, \frac{1}{x^2}\Big[ \left(1-x^2\right)^2 \left(
     1+4x^2\right)-2 \left(2-6x^2+2x^4-x^6\right) y^2
     +\left(6-12x^2-x^4\right) y^4 \nonumber \\
     && \hspace{1.2cm}-2 \left(2+x^2\right) y^6
     +y^8 \Big]\ln (\omega_1) \nonumber \\ [1.5mm]
     &&-4 \left(1-y^2\right) \Big[ 2+13x^2+x^4
     -\left(4+3x^2\right) y^2+2 y^4 \Big] \ln (\eta)
     \nonumber \\ [1.5mm]
     &&+\sqrt\lambda\Big[ 13+19x^2
     -\left(8-5x^2\right) y^2-5y^4 \Big] \bigg\}
  \\[12pt]
T_{ST^P} &=&-\, \frac{1}{\sqrt{2}x}\sqrt\lambda \left(
     1-x^2-y^2\right)
     +\frac{\alpha_sC_F}{4\pi} \times \frac{1}{\sqrt{2}x}\bigg\{
     4 \left(1-x^2+y^2\right) \left(1-x^2-y^2\right)
     {\cal{N}}_1 \nonumber \\ [1.5mm]
     &&+\frac{1}{x} \left[ (1-x)^2-y^2\right] \left[ 3(1-x)
     -(3-2x)y^2\right] {\cal{N}}_2 \nonumber \\ [1.5mm]
     &&-\frac{1}{x} \left[ (1+x)^2-y^2\right] \left[ 3(1+x)
     -(3+2x)y^2\right] {\cal{N}}_3 \nonumber \\ [1.5mm]
     &&-\frac{\sqrt\lambda}{x^2}\Big[ \left(1-x^2\right)
     \left(1+9x^2\right)
     -2 \left(1+10x^2\right) y^2+y^4 \Big]\ln (y)
     \nonumber \\ [1.5mm]
     &&-8\sqrt\lambda \left(1-x^2-y^2\right)
     \ln \left(\frac{x}{\lambda}\right) \nonumber \\ [1.5mm]
     &&+\, \frac{1}{2x^2}\Big[ \left(1-x^2\right)^2
     \left(1+9x^2\right)-3 \left(1-x^4\right) y^2
     +\left(3+5x^2\right) y^4-y^6 \Big]\ln (\omega_1)
     \nonumber \\ [1.5mm]
     &&-2\Big[ \left(2+x^2\right) \left(5+x^2\right)
     -\left(20+3x^2\right) y^2+10y^4 \Big]\ln (\eta)
     \nonumber \\ [1.5mm]
     &&+2\sqrt\lambda
     \left(5+4x^2-2y^2\right) \! \bigg\}
  \\[12pt]
T_{SN^P} &=& \frac{\lambda}{\sqrt{2}x}
     +\frac{\alpha_sC_F}{4\pi} \times \frac{1}{\sqrt{2}x}\bigg\{
     4\sqrt\lambda \left(1-x^2+y^2\right)
     {\cal{N}}_4 \nonumber \\ [1.5mm]
     &&+2\Big[ 1+7x^2-2x^4+\left(1+4 x^2\right) y^2
     -2y^4 \Big]{\cal{N}}_5
     \nonumber \\ [1.5mm]
     &&-\frac{1}{x^2}\lambda \left(1+9x^2-y^2\right) \ln (y)
     \nonumber \\ [1.5mm]
     &&+\frac{1}{2x^2}\sqrt\lambda
     \left[ 1+10x^2+x^4-2 \left(1+x^2\right) y^2+y^4\right]
     \ln (\omega_1)
     +8\lambda\ln \left[ (1-x)^2-y^2\right]
     \nonumber \\ [1.5mm]
     &&+2\Big[ \left(1-x^2\right) \left(1+5x^2\right)
     -\left(7-3x^2\right) y^2 \Big]
     \ln \left(\frac{1-x}{y}\right) \nonumber \\ [1.5mm]
     &&-2 \left[ (1-x)^2-y^2\right]
     \left(2-3x+4x^2-5y^2\right) \! \bigg\}
\end{eqnarray}
The dilogarithmic decay rate terms occuring in these expressions are
\begin{eqnarray}
{\cal{N}}_1 &=& 2 \left[ 2\Li_2 (1-\omega_1)
     -\Li_2 (\eta x)
     +\Li_2 \left(\frac{x}{\eta}\right)\right] \nonumber \\ [1.5mm]
     &&+2\ln (\omega_1)\ln (1-\omega_1)
     -\ln (\omega_1)\ln (x)-2\ln (\eta)\ln (y)
     \nonumber \\ [6mm]
{\cal{N}}_2 &=& \left[ 2\Li_2 (1-\omega_1)
     -\Li_2 (\eta x)
     +\Li_2 \left(\frac{x}{\eta}\right)\right]
     +2 \left[ \Li_2 \left(-\, \frac{\omega_1}{\eta}\right)
     -\Li_2 \left(-\, \frac{1}{\eta}\right)\right]
     \nonumber \\ [1.5mm]
     &&-2\ln (\eta)\ln (1-x)+2\ln (\omega_1)
     \ln \left(1+\frac{\omega_1}{\eta}\right) \nonumber \\[6mm]
{\cal{N}}_3 &=& \left[ 2\Li_2 (1-\omega_1)
     -\Li_2 (\eta x)
     +\Li_2 \left(\frac{x}{\eta}\right)\right]
     +2 \left[ \Li_2 \left(\frac{\omega_1}{\eta}\right)
     -\Li_2 \left(\frac{1}{\eta}\right)\right]
     \nonumber \\ [1.5mm]
     &&-2\ln (\eta)\ln (1+x)+2\ln (\omega_1)
     \ln \left(1-\frac{\omega_1}{\eta}\right) \nonumber \\[6mm]
{\cal{N}}_4 &=& \left[ 2\Li_2 (1-\omega_1)
     -\Li_2 (\eta x)
     +\Li_2 \left(\frac{x}{\eta}\right)\right]
     +2 \left[ \Li_2 \left(-\, \frac{\omega_1}{\eta}\right)
     -\Li_2 \left(-\, \frac{1}{\eta}\right)\right]
     \nonumber \\ [1.5mm]
     &&+2\ln (\omega_1)\ln (1-\omega_1)
     -\ln (\omega_1)\ln (\eta x) \nonumber \\ [6mm]
{\cal{N}}_5 &=& \Li_2 (\eta x)
     +\Li_2 \left(\frac{x}{\eta}\right)
     -2\Li_2 (x).
\end{eqnarray}
As an illustration of our results, in Figure~\ref{figphi} we plot the
dependence of the normalized decay rate $\hat\Gamma^-$ in dependence on the
cosine of the polar angle $\theta$ for different azimuthal angles $\phi$. The
polarization angle is chosen to be $\theta_P=\pi/2$, i.e.\ orthogonal to the
momentum of the $W$ boson. The plots show the typical enhancement in the
backwards direction of the charged lepton for $\phi=0$ while for $\phi=\pi$
the rate in this direction is nearly extinguished. The $O(\alpha_s)$
corrections reduce the rate uniformly by about $13\%$.

\section{Nonperturbative corrections}
Both bottom and charm quarks are enclosed in heavy hadrons. This is accounted
for by adding nonperturbative corrections to the decay width. The analysis
done here is based on Ref.~\cite{Manohar:1993qn} and a series of publications
of our working group on semileptonic decays of $B$ mesons~\cite{Ali:1978it,%
Korner:1989qb,Balk:1993sz} and $\Lambda_b$ baryons~\cite{Konig:1993ze,%
Korner:1994jh,Korner:1995my,Diaconu:1995mp,Balk:1997fg,Korner:1998nc,%
Korner:2000zn,Korner:2000hx}. Note that the results of
Ref.~\cite{Manohar:1993qn} have been independently checked by the authors of
Refs.~\cite{Blok:1993va,Colangelo:2020vhu}. The calculations are done in the
framework of the heavy quark effective theory (HQET) and the method of
operator product expansion (OPE) as applied to heavy hadron decays. As before,
the dynamics of the hadron-side transitions is embodied in the hadron tensor
$W^{\mu\nu}$. However, in this case this tensor is given by the absorptive
part of the current--current correlator,
\begin{equation}
\tilde W^{\mu \nu}=-\frac{1}{\pi}\mbox{Im}\left(\Pi^{\mu\nu}\right),
\end{equation}
where the correlator
\begin{equation}
\Pi^{\mu\nu}(q^2,q_0) = -i \langle \Lambda_b (p, s)|\int d^4xe^{-iqx}
  {\cal T}\{J^{\mu\dagger}(x)J^\nu(0)\}|\Lambda_b (p, s) \rangle
\end{equation}
can be written again in terms of five unpolarized and nine polarized invariant
structure functions. These can be calculated in HQET and read up to order
$O(1/m_b^2)$~\cite{Manohar:1993qn}
\begin{eqnarray}
\Pi_1 &=& \frac{1}{2\Delta_0}(m_b-v \cdot q)(1+X_b)
     +\frac{2m_b}{3}(K_b+G_b) \left(\frac{-1}{2\Delta_0}
     +\frac{q^2-(v \cdot q)^2}{\Delta_0^2}\right) \nonumber \\ [1.5mm]
      &&+\frac{m_b(K_b+G_b)}{2\Delta_0}
     -\frac{m_b^2G_b}{3\Delta_0^2}(m_b-v \cdot q)
    \nonumber\\[7pt]
\Pi_2 &=& \frac{m_b}{\Delta_0}(1+X_b)+\frac{2m_b}{3}(K_b+G_b)
     \left(\frac{1}{\Delta_0}+\frac{2m_b(v \cdot q)}{\Delta_0^2}
    \right)+\frac{m_b(K_b+G_b)}{\Delta_0} \nonumber \\ [1.5mm]
     &&+\frac{4m_b^2K_b(v \cdot q)}{3\Delta_0^2}
     +\frac{2m_b^3G_b}{3\Delta_0^2}\nonumber\\[7pt]
\Pi_3 &=& \frac{1}{2\Delta_0}(1+X_b)-\frac{2m_b}{3}
     (K_b+G_b)\frac{m_b-v \cdot q}{\Delta_0^2}
     +\frac{2m_b^2K_b}{3\Delta_0^2}
     -\frac{m_b^2G_b}{3\Delta_0^2}\nonumber\\[7pt]
\Pi_4 &=& \frac{4m_b}{3\Delta_0^2}(K_b+G_b)\nonumber\\[7pt]
\Pi_5 &=& \frac{-1}{2\Delta_0}(1+X_b)
     -\frac{2m_b}{3}(K_b+G_b)
     \frac{2m_b+v \cdot q}{\Delta_0^2}
     +\frac{m_b^2G_b}{3 \Delta_0^2}\nonumber\\[7pt]
\Pi^P_1 &=&-\, \frac{1+\epsilon_b}{2\Delta_0}
     -\frac{5m_b}{3\Delta_0^2}(v \cdot q)K_b
     +\frac{4m_b^2K_b}{3\Delta_0^3} \left(
     q^2-(v \cdot q)^2\right)\nonumber\\[7pt]
\Pi^P_2 &=& \frac{4m_b^2K_b}{3\Delta_0^2}\nonumber\\[7pt]
\Pi^P_3 &=& \frac{2m_bK_b}{3\Delta_0^2}\nonumber\\[7pt]
\Pi^P_4 &=& 0\nonumber\\[7pt]
\Pi^P_5 &=&-\, \frac{2m_bK_b}{3\Delta_0^2}\nonumber\\[7pt]
\Pi^P_6 &=&-\, \frac{m_b(1+\epsilon_b)}{2\Delta_0}
     -\frac{5m_bK_b}{6\Delta_0}
     -\frac{5m_b^2}{3\Delta_0^2}(v \cdot q)K_b
     +\frac{4m_b^3K_b}{3\Delta_0^3} \left(
     q^2-(v \cdot q)^2\right)\nonumber\\[7pt]
\Pi^P_7 &=& \frac{1+\epsilon_b}{2\Delta_0}
     +\frac{(2m_b+3v \cdot q)m_bK_b}{3\Delta_0^2}
     -\frac{4m_b^2K_b}{3\Delta_0^3} \left(
     q^2-(v \cdot q)^2\right)\nonumber\\[7pt]
\Pi^P_8 &=& \frac{m_b(1+\epsilon_b)}{2\Delta_0}
     +\frac{m_bK_b}{6\Delta_0}
     +\frac{5m_b^2}{3\Delta_0^2}(v \cdot q)K_b
     -\frac{4m_b^3K_b}{3\Delta_0^3} \left(
     q^2-(v \cdot q)^2\right)\nonumber\\[7pt]
\Pi^P_9 &=&-\, \frac{1+\epsilon_b}{2\Delta_0}
     -\frac{(2m_b+3v \cdot q)m_bK_b}{3\Delta_0^2}
     +\frac{4m_b^2K_b}{3\Delta_0^3} \left(
     q^2-(v \cdot q)^2\right)
\end{eqnarray}
where
\begin{equation}
X_b =-\, \frac{2(m_b-v \cdot q)m_b(K_b+G_b)}{\Delta_0}
     -\frac{8m_b^2K_b}{3\Delta_0^2} \left(
     q^2-(v \cdot q)^2\right)
     +\frac{2m_b^2K_b}{\Delta_0}
\end{equation}
and the denominator factor $\Delta_0$ is given by
\begin{equation}
\Delta_0 = (m_bv-q)^2-m_c^2+i\epsilon.
\end{equation}
$K_b$ is related to the mean kinetic energy of the heavy bottom quark
\begin{equation}\label{defKb}
K_b =-\sum_s \langle \Lambda_b (p, s) | \bar{b}_v (x_l)
     \frac{(i D)^2}{2m_b^2}b_v (x_l) | \Lambda_b (p, s) \rangle
      = \frac{\mu_{\pi}^2}{2m_b^2},
\end{equation}
where we can use $\mu_{\pi}^2 \approx 0.6\GeV^2$~\cite{Diaconu:1995mp}. The
spin dependent contribution $\epsilon_b$ is defined by
\begin{equation}\label{defyb}
\langle \Lambda_b (p, s) | \bar{b}\gamma^{\lambda}\gamma_5b |
     \Lambda_b (p, s) \rangle = (1+\epsilon_b)s^{\lambda}
\end{equation}
and is of the order $\Lambda_{QCD}^2/m_b^2$. Finally, the chromomagnetic
contribution $G_b$ is given by
\begin{equation}\label{defGb}
G_b = \sum_s \langle \Lambda_b (p, s) | \bar{b}_v (x_l) \left(
     \frac{-gF_{\alpha \beta}\sigma^{\alpha \beta}}{4m_b^2}
    \right) b_v (x_l) | \Lambda_b (p, s) \rangle = \frac{\mu_G^2}{2m_b^2}
\end{equation}
of the same order. Invariant structure functions $W_i$ are defined accordingly
by
\begin{equation}
\tilde W_i =-\frac1\pi\mbox{Im}(\Pi_i).
\end{equation}
The imaginary parts of the inverse powers of $\Delta_0$, which are needed
for obtaining $W_i$, can be calculated with the help of
\begin{eqnarray}
\mbox{Im} \left(\frac{1}{\Delta_0}\right) &=& \frac{- \pi}{2m_b}
     \delta \left[ q_0-\left(\frac{m_b^2-m_c^2+q^2}{2m_b}\right)
    \right]\nonumber\\[7pt]
\mbox{Im} \left(\frac{1}{\Delta_0^2}\right) &=& \frac{- \pi}{4m_b^2}
     \frac{d}{d q_0}\delta \left[ q_0
     -\left(\frac{m_b^2-m_c^2+q^2}{2m_b}\right)\right]
    \nonumber\\[7pt]
\mbox{Im} \left(\frac{1}{\Delta_0^3}\right) &=& \frac{- \pi}{16m_b^3}
     \frac{d^2}{d q_0^2}\delta \left[ q_0
     -\left(\frac{m_b^2-m_c^2+q^2}{2m_b}\right)\right].
\end{eqnarray}
Accordingly, the invariant structure functions are given by
\begin{equation}
\tilde W_i = \frac{1}{2m_b}\Pi_i
     \Bigg|\Bigg\{\frac{1}{\Delta_0}\rightarrow
     \delta (q_0-E_q), \frac{1}{\Delta_0^2}\rightarrow
     \frac{1}{2m_b}\delta^{\prime} (q_0-E_q),
     \frac{1}{\Delta_0^3}\rightarrow \frac{1}{8m_b^2}
     \delta^{\prime \prime} (q_0-E_q) \Bigg\}
\end{equation}
where $E_q = (m_b^2-m_c^2+q^2)/(2m_b)$.

The helicity structure functions $\tilde W_X$ can be obtained by linear
combinations of the invariant structure functions $\tilde W^{(P)}_i$ with the
help of Eqs.~(\ref{Wh}). By integrating over the $W$ energy scale $x_0$, one
obtains the integrated structure functions
\begin{equation} \label{IW}
T_X(x^2) := 8m_b \int \sqrt{x_0^2-x^2} \tilde W_X(x^2,x_0) dx_0
\end{equation}
which are given by
\begin{eqnarray}
T_U &=& 2(1-K_b)\sqrt\lambda(1-x^2+y^2)+\frac{16}{3}K_b\sqrt\lambda
     \nonumber \\ [1.5mm] &&
     +\frac{G_b}{3\sqrt\lambda}\Bigg\{2\lambda\Big[15(x^2-y^2)-11\Big]
     +8x^2\Big(3(x^2-y^2)-7\Big)+32(1-y^2)\Bigg\}\nonumber\\[7pt]
T_{U^P} &=&-2(1+\epsilon_b)\lambda+\frac23K_b(3\lambda+8x^2)\nonumber\\[7pt]
T_L &=& (1-K_b)\frac{\sqrt\lambda}{x^2}\Big[\lambda+x^2(1-x^2+y^2)\Big]
     -\frac{16}{3}K_b\sqrt\lambda\nonumber \\ [1.5mm] &&
     +\frac{G_b}{3\sqrt\lambda x^2}\Bigg\{\lambda
     \Big[15(-\lambda+x^4-x^2y^2)-59x^2+12(1-y^2)\Big]
     \nonumber \\ [1.5mm] && \hspace{2cm}
     +4x^2\Big[x^2\Big(3(x^2-y^2)-7\Big)+4(1-y^2)\Big] \Bigg\}\nonumber\\[7pt]
T_{L^P} &=& (1+\epsilon_b)\frac{\lambda}{x^2}(1-y^2)
     -K_b\frac{1-y^2}{3x^2}(3\lambda+8x^2)\nonumber\\[7pt]
T_S &=& (1-K_b)\frac{\sqrt\lambda}{x^2}\Big[\lambda+x^2(1-x^2+y^2)\Big]
     \nonumber \\ [1.5mm] &&
     +\frac{G_b}{\sqrt\lambda x^2}\Bigg\{\lambda
     \Big[-5(\lambda-x^4+x^2y^2)-9x^2+4(1-y^2)\Big]
     -4x^4(1-x^2+y^2)\Bigg\}\nonumber\\[7pt]
T_{S^P} &=& (1+\epsilon_b)\frac{\lambda}{x^2}(1-y^2)
     -K_b\frac{1-y^2}{3x^2}(3\lambda+8x^2)\nonumber\\[7pt]
T_F &=&-2\lambda+\frac23K_b(3\lambda+8x^2)
     +\frac23G_b\left[15\lambda+16x^2-24(1-y^2)\right]\nonumber\\[7pt]
T_{F^P} &=& 2(1+\epsilon_b)\sqrt\lambda(1-x^2+y^2)
     +\frac23K_b\sqrt\lambda\Big(3(x^2-y^2)+5\Big)\nonumber\\[7pt]
T_{I^P} &=&-(1+\epsilon_b)\frac{\lambda}{\sqrt{2}x}
     -2K_b\frac{\lambda-4x^2}{3\sqrt{2}x}\nonumber\\[7pt]
T_{A^P} &=& (1+\epsilon_b)\frac{\sqrt\lambda}{\sqrt{2}x}(1-x^2-y^2)
     -2K_b\frac{\sqrt\lambda}{3\sqrt{2}x}(1+x^2-y^2)\nonumber\\[7pt]
T_{SL} &=& \frac{\lambda}{x^2}(1-y^2)-K_b\frac{1-y^2}{3x^2}(3\lambda+8x^2)
     -G_b\frac{1-5y^2}{3x^2}(3\lambda+8x^2)\nonumber\\[7pt]
T_{SL^P} &=& (1+\epsilon_b)\frac{\sqrt\lambda}{x^2}
     \Big[\lambda+x^2(1-x^2+y^2)\Big]-K_b\frac{\sqrt\lambda}{3x^2}
     \Big[3\lambda-x^2\Big(3(x^2-y^2)-11\Big)\Big]\nonumber\\[7pt]
T_{ST^P} &=&-(1+\epsilon_b)\frac{\sqrt\lambda}{\sqrt{2}x}(1-x^2-y^2)
     -2K_b\frac{\sqrt\lambda}{3\sqrt{2}x}(1+x^2-y^2)\nonumber\\[7pt]
T_{SN^P} &=& (1+\epsilon_b)\frac{\lambda}{\sqrt{2}x}
     -2K_b\frac{\lambda+4x^2}{3\sqrt{2}x},
\end{eqnarray}
where $K_b$, $\epsilon_b$ and $G_b$ are defined in Eqs.~(\ref{defKb}),
(\ref{defyb}) and~(\ref{defGb}), respectively. Taking typical values for these
parameters advised for instance in Refs.~\cite{Manohar:1993qn,Diaconu:1995mp},
the corrections to Fig.~\ref{figphi} are below the $1\%$ level and, therefore,
outplayed by the first order radiative QCD corrections.

\begin{figure}
\epsfig{figure=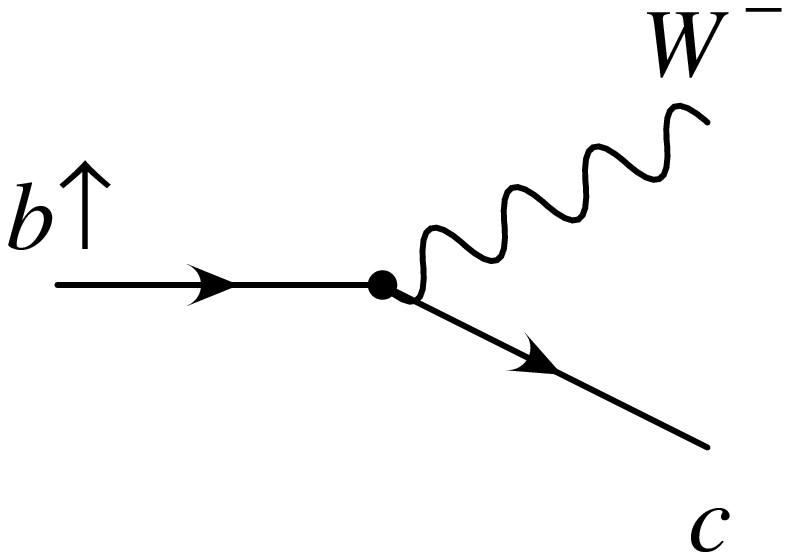, scale=0.47}\quad
\epsfig{figure=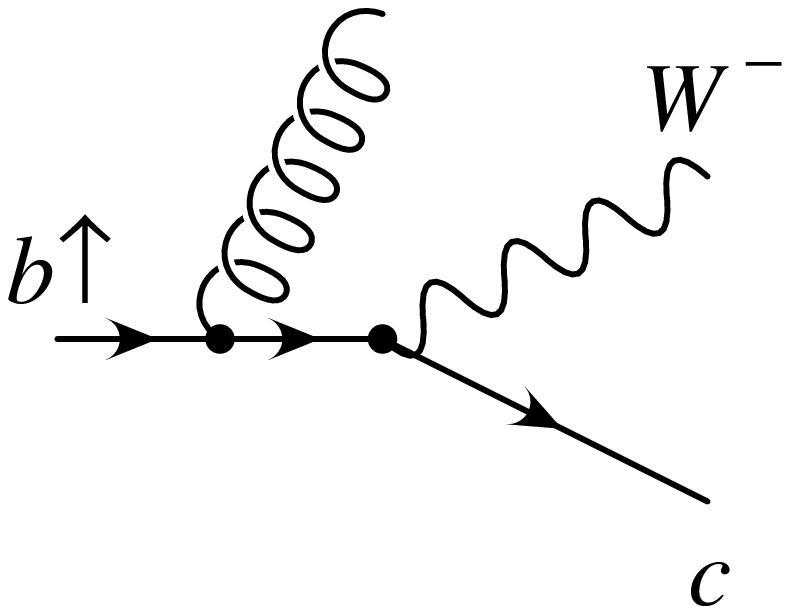, scale=0.47}\quad
\epsfig{figure=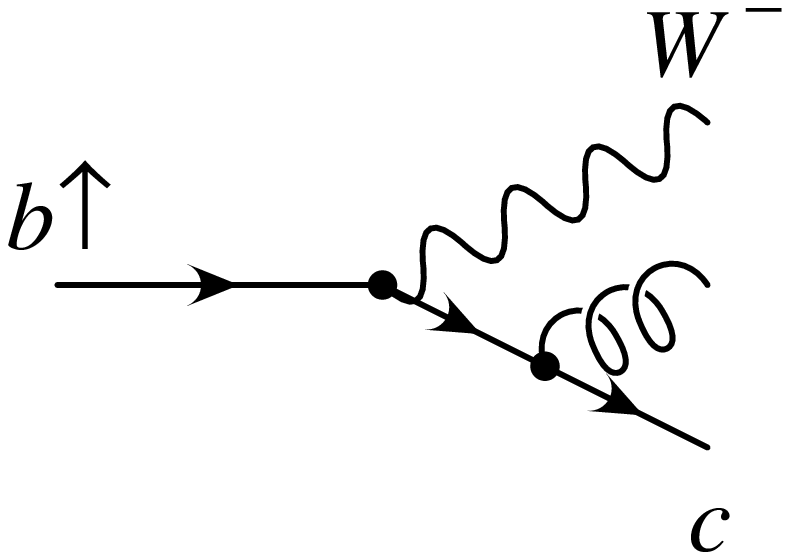, scale=0.47}\quad
\epsfig{figure=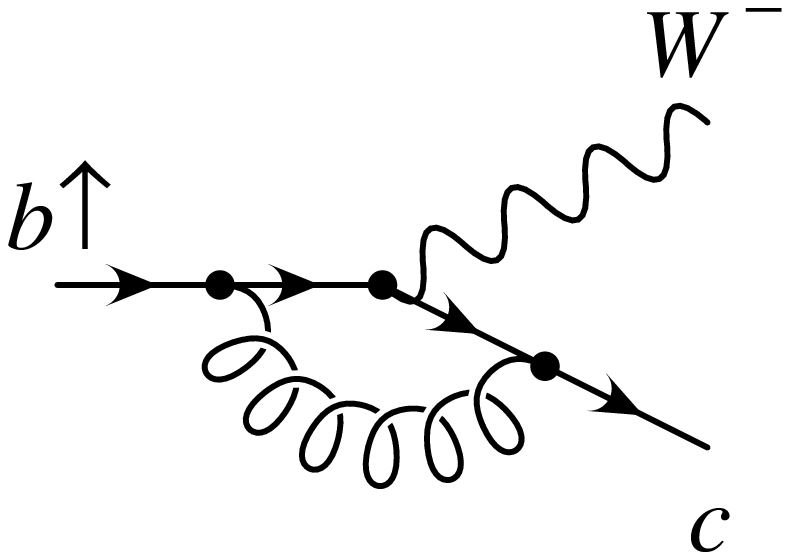, scale=0.47}
\caption{\label{diagrams}Feynman diagrams for Born term, QCD first order tree
and loop contributions}
\end{figure}

\section{Summary}
In this paper we have given analytical expressions for first order radiative
QCD corrections (cf.\ Fig.~\ref{diagrams}) to the the ten helicity structure
functions that determine the angular decay distribution of the semileptonic
decay of a polarized bottom quark. We have shown that the radiative
corrections change the Born term result significantly. At the same time, we
have found that nonperturbative corrections calculated in some detail in the
last part of this paper are subdominant.

\subsection*{Acknowledgment}
We gratefully remember our deceased collaborator J\"urgen G.~K\"orner who was
one of the initiators of this project. The research was supported in part by
the European Regional Development Fund under Grant No.~TK133. S.G. acknowledges
continuing support by the PRISMA/PRISMA+ Cluster of Excellence at the
Johannes Gutenberg University in Mainz which allowed to finish this project
during his stays in Mainz.

\begin{appendix}

\section{Integrated helicity rates}
\setcounter{equation}{0}\def\theequation{A\arabic{equation}}
We integrate the analytical results for the helicity rates
$d\hat{\Gamma}_i^{(incl)}/dx^2$ defined in (\ref{gamma})
over the scaled $W$ boson momentum squared $x^2$. The limits of integration
in case of massive leptons are $\zeta^2 \leq x^2 \leq (1-y)^2$, and the
analytical results for the integrated helicity rates are given by
\begin{equation}
\hat{\Gamma} = \int_{\zeta^2}^{(1-y)^2}\frac{d\hat{\Gamma}}{dx^2}dx^2.
\end{equation}
With the definitions
\begin{equation}
u_1 = \frac{1-\zeta^2+y^2-\sqrt{R}}{2y},\qquad
u_2 = \frac{1+\zeta^2-y^2-\sqrt{R}}{2\zeta},\qquad
u_3 = \frac{(1-y)^2}{\zeta^2}
\end{equation}
we obtain the integrated rates into negative helicity leptons
\begin{eqnarray} \label{gammaminus}
\hat{\Gamma}_U^- &=& \frac{1}{3}\sqrt{R}\Big[
     \left(1+y^2\right)
     \left(1-8y^2+y^4\right)
     -\left(7-12y^2+7y^4
    \right) \zeta^2-7 \left(1+y^2\right) \zeta^4+\zeta^6 \Big]
     \nonumber \\ [1.5mm]
     &&-8y^4 \left(1-\zeta^4\right) \ln (u_1)
     -8 \left(1-y^4\right) \zeta^4\ln (u_2)
     \nonumber \\ [1.5mm]
     &+& K_b\Bigg\{ \frac{1}{9}\sqrt{R}\Big[
     13+181y^2+37y^4-3y^6
     \nonumber \\ [1.5mm]
     && \hspace{1.5cm}-\left(59+116y^2-21y^4
    \right) \zeta^2-\left(11-21y^2\right) \zeta^4
     -3\zeta^6 \Big] \nonumber \\ [1.5mm]
     &&+\frac{8}{3}y^2\Big[ 8+11y^2
    -16\zeta^2+\left(8-3y^2\right) \zeta^4 \Big]
     \ln (u_1) \nonumber \\ [1.5mm]
     &&-\frac{8}{3} \left(1-y^2\right)
     \left(5-3y^2\right)
     \zeta^4\ln (u_2) \Bigg\} \nonumber \\ [1.5mm]
     &+& G_b\Bigg\{ \frac{1}{9}\sqrt{R}\Big[
     41+185y^2+65y^4-15y^6
     \nonumber \\ [1.5mm]
     && \hspace{1.5cm}-\left(127+124y^2
     -105y^4 \right) \zeta^2
     +\left(41+105y^2\right) \zeta^4-15\zeta^6 \Big]
     \nonumber \\ [1.5mm]
     &&+\frac{8}{3}y^2\Big[ 12+11y^2
     -16\zeta^2+\left(4-15y^2\right) \zeta^4 \Big]
     \ln (u_1) \nonumber \\ [1.5mm]
     &&-\frac{8}{3} \left(5-4y^2+15y^4\right)
     \zeta^4\ln (u_2) \Bigg\}\nonumber\\[7pt]
\hat{\Gamma}_{U^P}^- &=& (1+\epsilon_b)\Bigg\{ \frac{1}{3}
     \Big[ \zeta^2-(1-y)^2 \Big] \Big[
     (1-y)^4 \left(1+6y+y^2\right)
     \nonumber \\ [1.5mm]
     && \hspace{1.5cm}-(1-y)^2 \left(7+26y
     +7y^2\right) \zeta^2-\left(7+2y
     +7y^2\right) \zeta^4+\zeta^6 \Big] \nonumber \\ [1.5mm]
     \noalign{\vfill}
     &&-4 \left(1-y^2\right)^2\zeta^4\ln (u_3) \Bigg\}
     \nonumber \\ [1.5mm]
     \noalign{\vfill}
     &-& K_b\Bigg\{ \frac{1}{9}\Big[ \zeta^2-(1-y)^2
     \Big]\Big[ (1-y)^4 \left(35+18y
     +3y^2\right) \nonumber \\ [1.5mm]
     \noalign{\vfill}
     && \hspace{1.5cm}-(1-y)^2 \left(85+78y
     +21y^2\right) \zeta^2+\left(11-6y
     -21y^2\right) \zeta^4+3\zeta^6 \Big]
     \nonumber \\ [1.5mm]
     &&-4 \left(1-y^2\right)^2 \zeta^4\ln (u_3) \Bigg\}
    \nonumber\\[7pt]
     \noalign{\vfill}
\hat{\Gamma}_L^- &=& \frac{2}{3}\sqrt{R}\Big[
     \left(1+y^2\right)
     \left(1-8y^2+y^4\right)
     +10 \left(1-y^2+y^4\right)
     \zeta^2+\left(1+y^2\right) \zeta^4 \Big] \nonumber \\ [1.5mm]
     &&- 8y^4\Big[ 2-\left(3-y^2\right)
     \zeta^2+\zeta^4 \Big]\ln (u_1) \nonumber \\ [1.5mm]
     &&+8\zeta^2 \left(1-y^2\right)
     \Big[ \left(1-y^2\right)^2
     +\left(1+y^2\right) \zeta^2 \Big]\ln (u_2)
     \nonumber \\ [1.5mm]
     &+& K_b\Bigg\{-\, \frac{2}{9}\sqrt{R}\Big[ 11
     +59y^2-13y^4+3y^6
     \nonumber \\ [1.5mm]
     && \hspace{1.5cm}-10 \left(1+7y^2
     -3y^4\right) \zeta^2-\left(13-3y^2\right)
     \zeta^4 \Big] \nonumber \\ [1.5mm]
     \noalign{\vfill}
     &&-\frac{8}{3}y^2\Big[ 2 \left(4+y^2\right)
     -\left(16-9y^2+3y^4\right) \zeta^2
     +\left(8-3y^2\right) \zeta^4 \Big]\ln (u_1)
     \nonumber \\ [1.5mm]
     &&-\frac{8}{3}\zeta^2 \left(1-y^2\right) \Big[
     3 \left(1-y^2\right)^2-\left(5-3y^2\right)
     \zeta^2 \Big]\ln (u_2) \Bigg\} \nonumber \\ [1.5mm]
     \noalign{\vfill}
     &+& G_b\Bigg\{-\, \frac{2}{9}\sqrt{R}\Big[ 7
     +115y^2-53y^4+15y^6
     \nonumber \\ [1.5mm]
     && \hspace{1.5cm}-2 \left(1+97y^2
     -75y^4\right) \zeta^2
     -\left(17-15y^2\right) \zeta^4 \Big] \nonumber \\ [1.5mm]
     &&-\frac{8}{3}y^2\Big[ 2 \left(6+y^2\right)
     -\left(28-33y^2+15y^4 \right) \zeta^2
     +\left(16-15y^2\right) \zeta^4 \Big]\ln (u_1)
     \nonumber \\ [1.5mm]
     \noalign{\vfill}
     &&-\frac{8}{3}\zeta^2 \Big[ 3 \left(1-y^2\right)^2
     \left(1-5y^2\right)
     -\left(5-16y^2+15y^4\right)
     \zeta^2 \Big]\ln (u_2) \Bigg\}\nonumber\\[7pt]
\hat{\Gamma}_{L^P}^- &=& (1+\epsilon_b)\Bigg\{ \frac{2}{3}
     \left(1-y^2\right) \Big[ (1-y)^2-\zeta^2 \Big]
     \nonumber \\ [1.5mm]
     && \hspace{1.5cm} \times \Big[ (1-y)^2
     \left(1+4y+y^2\right)
     +10 \left(1+y+y^2\right) \zeta^2+\zeta^4 \Big]
     \nonumber \\ [1.5mm]
     &&-4 \left(1-y^2\right) \zeta^2\Big[
     \left(1-y^2\right)^2
     +\left(1+y^2\right) \zeta^2 \Big]\ln (u_3) \Bigg\}
     \nonumber \\ [1.5mm]
     &-& K_b\Bigg\{ \frac{2}{3} \left(1-y^2\right) \Big[
     (1-y)^2-\zeta^2 \Big] \nonumber \\ [1.5mm]
     && \hspace{1.5cm} \times \Big[ (1-y)^2
     \left(5+4y+y^2\right)
     -2 \left(1-5y-5y^2\right) \zeta^2
     +\zeta^4 \Big] \nonumber \\ [1.5mm]
     &&-\frac{4}{3} \left(1-y^2\right) \zeta^2
     \Big[ 3 \left(1-y^2\right)^2
     -\left(1-3y^2\right) \zeta^2 \Big]\ln (u_3) \Bigg\}
    \nonumber\\[7pt]
\hat{\Gamma}_F^- &=&-\, \frac{1}{3}\Big[ (1-y)^2
     -\zeta^2 \Big]\Big[ (1-y)^4
     \left(1+6y+y^2\right) \nonumber \\ [1.5mm]
     && \hspace{1.5cm}-(1-y)^2
     \left(7+26y+7y^2\right)
     \zeta^2-\left(7+2y+7y^2\right) \zeta^4
     +\zeta^6 \Big] \nonumber \\ [1.5mm]
     &&-4 \left(1-y^2\right)^2\zeta^4\ln (u_3)
     \nonumber \\ [1.5mm]
     &+& K_b\Bigg\{ \frac{1}{9}\Big[ (1-y)^2
     -\zeta^2 \Big]\Big[ (1-y)^4
     \left(35+18y+3y^2\right) \nonumber \\ [1.5mm]
     && \hspace{1.5cm}-(1-y)^2
     \left(85+78y+21y^2\right) 
     \zeta^2+\left(11-6y-21y^2\right) \zeta^4
     +3\zeta^6 \Big] \nonumber \\ [1.5mm]
     \noalign{\vfill}
     &&+4 \left(1-y^2\right)^2 \zeta^4\ln (u_3) \Bigg\}
     \nonumber \\ [1.5mm]
     \noalign{\vfill}
     &+& G_b\Bigg\{-\, \frac{1}{9}\Big[ (1-y)^2
     -\zeta^2 \Big] \Big[ (1-y)^3
     \left(65+133y+75y^2+15y^3\right)
     \nonumber \\ [1.5mm]
     && \hspace{1.5cm}-(1-y)
     \left(199+275y+285y^2
     +105y^3\right) \zeta^2 \nonumber \\ [1.5mm]
     \noalign{\vfill}
     && \hspace{1.5cm}+\left(41
     +30y+105y^2\right) \zeta^4
     -15\zeta^6 \Big] \nonumber \\ [1.5mm]
     &&-4 \left(1-y^2\right)
     \left(3+5y^2\right) \zeta^4\ln (u_3) \Bigg\}
    \nonumber\\[7pt]
\hat{\Gamma}_{F^P}^- &=& (1+\epsilon_b)\Bigg\{ \frac{1}{3}
     \sqrt{R}\Big[ \left(1+y^2\right)
     \left(1-8y^2+y^4\right) \nonumber \\ [1.5mm]
      \noalign{\vfill}
     && \hspace{1.5cm}-\left(7-12y^2
     +7y^4\right) \zeta^2-7 \left(1+y^2\right)
     \zeta^4+\zeta^6 \Big] \nonumber \\ [1.5mm]
     &&-8y^4 \left(1-\zeta^4\right) \ln (u_1)
     -8\zeta^4 \left(1-y^4\right) \ln (u_2) \Bigg\}
     \nonumber \\ [1.5mm]
     &+& K_b\Bigg\{ \frac{1}{9}\sqrt{R}\Big[
     13+181y^2+37y^4-3y^6
     \nonumber \\ [1.5mm]
     \noalign{\vfill}
     && \hspace{1.5cm}-\left(59+116y^2
     -21y^4\right) \zeta^2
     -\left(11-21y^2\right) \zeta^4
     -3\zeta^6 \Big] \nonumber \\ [1.5mm]
     &&+\frac{8}{3}y^2
     \Big[ 8+11y^2-16\zeta^2
     +\left(8-3y^2\right) \zeta^4 \Big]\ln (u_1)
     \nonumber \\ [1.5mm]
     &&-\frac{8}{3}\zeta^4 \left(1-y^2\right)
     \left(5-3y^2\right) \ln (u_2) \Bigg\}\nonumber\\[7pt]
\hat{\Gamma}_{I^P}^- &=&-\, (1+\epsilon_b)
     \frac{16\sqrt{2}}{105}(1-y-\zeta)^4\Big[
     (1-y) \left(1+5y+y^2\right)
     \nonumber \\ [1.5mm]
     && \hspace{1.5cm}+4 \left(1+5y
     +y^2\right) \zeta
     -4(1-y)\zeta^2-\zeta^3 \Big] \nonumber \\ [1.5mm]
     &+& K_b\frac{16\sqrt{2}}{315}(1-y-\zeta)^3
     \Big[ (1-y)^2 \left(19-10y
     -2y^2\right) \nonumber \\ [1.5mm]
     && \hspace{1.5cm}+3(1-y)
     \left(19-10y-2y^2
    \right) \zeta+8 \left(9+3y+2y^2\right)
     \zeta^2 \nonumber \\ [1.5mm]
     && \hspace{1.5cm}-6(1-y)\zeta^3-2\zeta^4 \Big]
    \nonumber\\[7pt]
\hat{\Gamma}_{A^P}^- &=& (1+\epsilon_b) \Bigg\{
     \frac{8\sqrt{2}}{105}\zeta\sqrt{R}\Big[ 
     2-13y^2-5y^4+11
     \left(4-3y^2\right) \zeta^2+2\zeta^4 \Big]
     \nonumber \\ [1.5mm]
     &&+\frac{8\sqrt{2}}{105}(1+y)
     \Big[ 2-17y^2-108y^4-5y^6
     \nonumber \\ [1.5mm]
     && \hspace{1.5cm}-14 \left(2-7y^2-3
     y^4\right) \zeta^2-35 \left(2-y^2\right)
     \zeta^4 \Big]\Big[ \mbox{E} \! \left(k^2\right)-\mbox{E} \!
     \left(\varphi, k^2\right) \Big] \nonumber \\ [1.5mm]
     &&-\frac{16\sqrt{2}}{105}y\Big[ (1+y)
     \left(2-3y-13y^2-45y^3-5
     y^4\right) \nonumber \\ [1.5mm]
     && \hspace{1.5cm}-14(1+y)
     \left(2-3y-3y^2\right)
     \zeta^2-35 \left(2-y\right) \zeta^4 \Big]\Big[
     \mbox{K} \! \left(k^2\right)-\mbox{F} \! \left(\varphi, k^2\right)
     \Big] \Bigg\} \nonumber \\ [1.5mm]
     &-& K_b\Bigg\{ \frac{16\sqrt{2}}{315}\zeta\sqrt{R}
     \Big[ 5+13y^2-2y^4+11
     \left(3-4y^2\right) \zeta^2
    -2\zeta^4 \Big] \nonumber \\ [1.5mm]
     &&+\frac{16\sqrt{2}}{315}(1+y)
     \Big[ 5+108y^2+17y^4-2y^6
     \nonumber \\ [1.5mm]
     && \hspace{1.5cm}-14(3+7y^2
     -2y^4)\zeta^2-35 \left(1-2y^2\right)
     \zeta^4 \Big]\Big[ \mbox{E} \! \left(k^2\right)-\mbox{E} \!
     \left(\varphi, k^2\right) \Big] \nonumber \\ [1.5mm]
     &&-\frac{32\sqrt{2}}{315}y \Big[ (1+y) \left(
     5+45y+13y^2+3y^3
     -2y^4\right) \nonumber \\ [1.5mm]
     && \hspace{1.5cm}-14(1+y)
     \left(3+3y-2y^2\right)
     \zeta^2-35 \left(1- 2y\right) \zeta^4 \Big]
     \Big[ \mbox{K} \! \left(k^2\right)-\mbox{F} \!
     \left(\varphi, k^2\right) \Big] \Bigg\}. \nonumber \\
\end{eqnarray}
The integrated rates into positive helicity leptons read
\begin{eqnarray} \label{gammaplus}
\hat{\Gamma}_U^+ &=& \frac{2}{3}\sqrt{R}\zeta^2\Big[ \left(
     1-y^2\right)^2+10 \left(1+y^2\right) \zeta^2
     +\zeta^4 \Big] \\ [1.5mm]
     &&-8\frac{y^4\zeta^4}{1-y^2}
     \left(1-y^2-\zeta^2
    \right) \ln (u_1) \nonumber \\ [1.5mm]
     \noalign{\vfill}
     &&+8\frac{\zeta^4}{1-y^2}
     \Big[ \left(1-y^2\right)^2
     \left(1+y^2\right)+\left(1+y^4\right) \zeta^2
     \Big] \ln (u_2) \nonumber \\ [1.5mm]
     &+& K_b\Bigg\{ \frac{2}{3}\sqrt{R}\zeta^2\Big[
     3+6y^2-y^4+10
     \left(1-y^2\right) \zeta^2-\zeta^4 \Big] \nonumber \\ [1.5mm]
     &&+\frac{8}{3}\,\frac{y^2\zeta^2}{1-y^2}
     \Big[ 4 \left(1-y^2\right)-\left(1-y^2\right)
     \left(8-3y^2\right) \zeta^2
     +\left(4-3y^2\right) \zeta^4\Big] \ln (u_1)
     \nonumber \\ [1.5mm]
     &&+\frac{8}{3}\,\frac{\zeta^4}{1-y^2}
     \Big[ \left(1-y^2\right)^2 \left(5-3y^2\right)
     +\left(1+4y^2-3y^4
    \right) \zeta^2 \Big]\ln (u_2) \Bigg\} \nonumber \\ [1.5mm]
     &+& G_b\Bigg\{ \frac{2}{3}\sqrt{R}
     \frac{\zeta^2}{1-y^2}\Big[ \left(1-y^2\right)
     \left(7+6y^2-5y^4\right) \nonumber \\ [1.5mm]
     && \hspace{1.5cm}-2 \left(3+14y^2
     -25y^4\right) \zeta^2
     -5 \left(1-y^2\right) \zeta^4 \Big] \nonumber \\ [1.5mm]
     &&+\frac{8}{3}
     \frac{y^2\zeta^2}{\left(1-y^2\right)^2}
     \Big[ 4 \left(1-y^2\right)^2-\left(1-y^2\right)^2
     \left(4-15y^2\right) \zeta^2 \nonumber \\ [1.5mm]
     \noalign{\vfill}
     && \hspace{1.5cm}-y^2 \left(23-15y^2\right)
     \zeta^4 \Big]\ln (u_1) \nonumber \\ [1.5mm]
     \noalign{\vfill}
     &&+\frac{8}{3}\,\frac{\zeta^4}{\left(1-y^2\right)^2}
     \Big[ \left(1-y^2\right)^2
     \left(5-4y^2+15y^4\right) \nonumber \\ [1.5mm]
     && \hspace{1.5cm}-\left(7+y^2+23
     y^4-15y^6\right) \zeta^2 \Big]\ln (u_2) \Bigg\}
    \nonumber\\[7pt]
\hat{\Gamma}_{U^P}^+ &=& (1+\epsilon_b)\Bigg\{ \frac{2}{3}
     \zeta^2\Big[ \zeta^2-(1-y)^2 \Big] \nonumber \\ [1.5mm]
     \noalign{\vfill}
     && \hspace{1.5cm} \times \Big[ (1-y)^2 \left(
     1+4y+y^2\right)+10
     \left(1+y+y^2\right) \zeta^2+\zeta^4
     \Big] \nonumber \\ [1.5mm]
     &&+4\zeta^4\Big[ \left(1-y^2\right)^2
     +\left(1+y^2\right) \zeta^2 \Big]\ln (u_3) \Bigg\}
     \nonumber \\ [1.5mm]
     &-& K_b\Bigg\{ \frac{2}{3}\zeta^2
     \Big[ \zeta^2-(1-y)^2
     \Big]\Big[ (1-y)^2
     \left(5+4y+y^2\right)
     -2 \left(1-5y-5y^2\right) \zeta^2
     +\zeta^4 \Big] \nonumber \\ [1.5mm]
     &&+\frac{4}{3}\zeta^4\Big[ 3 \left(1-y^2\right)^2
     -\left(1-3y^2\right) \zeta^2 \Big]\ln (u_3) \Bigg\}
    \nonumber\\[7pt]
     \noalign{\vfill}
\hat{\Gamma}_L^+ &=&-\, \sqrt{R}\zeta^2\Big[ 3-4y^2
     +3y^4+3 \left(1+y^2\right) \zeta^2 \Big]
     \nonumber \\ [1.5mm]
     &&-2\frac{y^4\zeta^2}{1-y^2}
     \Big[ \left(1-y^2\right) \left(3-y^2\right)
     -4 \left(1-y^2\right) \zeta^2
     +\zeta^4 \Big]\ln (u_1) \nonumber \\ [1.5mm]
     \noalign{\vfill}
     &&-2\frac{\zeta^2}{1-y^2}
     \Big[ \left(1-y^2\right)^4
     +4 \left(1-y^2\right)^2 \left(1+y^2\right) \zeta^2
     +\left(1+y^4\right) \zeta^4 \Big]\ln (u_2)
     \nonumber \\ [1.5mm]
     &+& K_b\Bigg\{ \frac{1}{3}\sqrt{R}\zeta^2 \Big[
     1-20y^2+9y^4
     -\left(31-9y^2\right) \zeta^2 \Big] \nonumber \\ [1.5mm]
     &&-\frac{2}{3}\,\frac{y^2\zeta^2}{1-y^2}
     \Big[ \left(1-y^2\right)
     \left(16-9y^2+3y^4\right) \nonumber \\ [1.5mm]
     && \hspace{1.5cm}-4 \left(1-y^2\right)
     \left(8-3y^2\right) \zeta^2
     +\left(16-3y^2\right) \zeta^4 \Big]\ln (u_1)
     \nonumber \\ [1.5mm]
     &&+\frac{2}{3}\,\frac{\zeta^2}{1-y^2}\Big[ 3 \left(
     1-y^2\right)^4-4 \left(1-y^2\right)^2
     \left(5-3y^2\right) \zeta^2 \nonumber \\ [1.5mm]
     && \hspace{1.5cm}-\left(13+16y^2-3
     y^4\right) \zeta^4 \Big]\ln (u_2) \Bigg\} \nonumber \\ [1.5mm]
     &+& G_b\Bigg\{ \frac{1}{3}\sqrt{R}
     \frac{\zeta^2}{1-y^2}\Big[ \left(1-y^2\right)
     \left(3-58y^2+45y^4\right)
     -\left(37-86y^2+45y^4\right) \zeta^2 \Big]
     \nonumber \\ [1.5mm]
     \noalign{\vfill}
     &&-\frac{2}{3}\,\frac{y^2\zeta^2}{\left(1
     -y^2\right)^2}\Big[ \left(1-y^2\right)^2
     \left(28-33y^2+15y^4\right)
     \nonumber \\ [1.5mm]
     && \hspace{1.5cm}-4 \left(1-y^2\right)^2 \left(
     16-15y^2\right) \zeta^2
     +\left(36-47y^2+15y^4
    \right) \zeta^4 \Big]\ln (u_1) \nonumber \\ [1.5mm]
     &&+\frac{2}{3}\,\frac{\zeta^2}{\left(1-y^2\right)^2}
     \Big[ 3 \left(1-y^2\right)^4 \left(1-5y^2\right)
    -4 \left(1-y^2\right)^2 \left(
     5-16y^2+15y^4\right) \zeta^2
     \nonumber \\ [1.5mm]
     && \hspace{1.5cm}-\left(17+23y^2
     -47y^4+15y^6\right) \zeta^4 \Big]
     \ln (u_2) \Bigg\}\nonumber\\[7pt]
\hat{\Gamma}_{L^P}^+ &=& (1+\epsilon_b)\Bigg\{ \zeta^2
      \frac{1+y}{1-y}\Big[ \zeta^2-(1-y)^2
     \Big] \nonumber \\ [1.5mm]
     && \hspace{1.5cm} \times \Big[ (1-y)^2
     \left(3+4y+3y^2\right)
     +\left(3-4y+3y^2\right) \zeta^2 \Big]
     \nonumber \\ [1.5mm]
     &&+\left(1-y^2\right) \zeta^2
     \Big[ \left(1-y^2\right)^2 +4 \left(1+y^2\right)
     \zeta^2+\zeta^4 \Big]\ln (u_3) \Bigg\} \nonumber \\ [1.5mm]
     &-& K_b \Bigg\{ \frac{1}{3}\zeta^2
     \frac{1+y}{1-y}\Big[ \zeta^2
     -(1-y)^2 \Big] \nonumber \\ [1.5mm]
     && \hspace{1.5cm} \times \Big[ (1-y)^2
     \left(1+12y+9y^2\right)
     +\left(1-12y+9y^2\right) \zeta^2 \Big]
     \nonumber \\ [1.5mm]
     &&+\frac{1}{3} \left(1-y^2\right) \zeta^2 \Big[ 3
     \left(1-y^2\right)^2
     -4 \left(1-3y^2\right) \zeta^2+3\zeta^4 \Big]
     \ln (u_3) \Bigg\}\nonumber\\[7pt]
\hat{\Gamma}_S^+ &=&-\, \sqrt{R}\zeta^2
     \Big[ 3-4y^2+3y^4
     +3 \left(1+y^2\right) \zeta^2 \Big] \nonumber \\ [1.5mm]
     \noalign{\vfill}
     &&-2\frac{y^4\zeta^2}{1-y^2}
     \Big[ \left(1-y^2\right)
     \left(3-y^2\right)-4 \left(1-y^2\right) \zeta^2
     +\zeta^4 \Big]\ln (u_1) \nonumber \\ [1.5mm]
     &&-2\frac{\zeta^2}{1-y^2}\Big[ \left(1-y^2
    \right)^4+4 \left(1-y^2\right)^2
     \left(1+y^2\right) \zeta^2
     +\left(1+y^4\right) \zeta^4 \Big]\ln (u_2)
     \nonumber \\ [1.5mm]
     &+& K_b\Bigg\{ \sqrt{R}\zeta^2\Big[ 3
     -4y^2+3y^4
     +3 \left(1+y^2\right) \zeta^2 \Big] \nonumber \\ [1.5mm]
     \noalign{\vfill}
     &&+2\frac{y^4\zeta^2}{1-y^2}
     \Big[ \left(1-y^2
    \right) \left(3-y^2\right)-4 \left(1-y^2\right)
     \zeta^2+\zeta^4 \Big]\ln (u_1) \nonumber \\ [1.5mm]
     \noalign{\vfill}
     &&+2\frac{\zeta^2}{1-y^2}
     \Big[ \left(1-y^2\right)^4
     +4 \left(1-y^2\right)^2 \left(1+y^2\right) \zeta^2
     +\left(1+y^4\right) \zeta^4 \Big]\ln (u_2) \Bigg\}
     \nonumber \\ [1.5mm]
     &+& G_b\Bigg\{ \sqrt{R}\,\frac{\zeta^2}{1-y^2}\Big[
     \left(1-y^2\right) \left(1-14y^2+15
     y^4\right)+\left(9+2y^2-15
     y^4\right) \zeta^2 \Big] \nonumber \\ [1.5mm]
     &&-2\frac{y^2\zeta^2}{\left(1
     -y^2\right)^2}\Big[ \left(1-y^2\right)^2
     \left(4-11y^2+5y^4\right) \nonumber \\ [1.5mm]
     && \hspace{1.5cm}+20y^2
     \left(1-y^2\right)^2 \zeta^2
     -\left(4+5y^2-5y^4\right) \zeta^4 \Big]
     \ln (u_1) \nonumber \\ [1.5mm]
     &&+2\frac{\zeta^2}{\left(1-y^2\right)^2}\Big[
     \left(1-y^2\right)^4 \left(1-5y^2\right)
     \nonumber \\ [1.5mm]
     && \hspace{1.5cm}+4 \left(1-y^2\right)^2
     \left(1-5y^4\right) \zeta^2
     +\left(5+3y^2+5y^4-5y^6\right)
     \zeta^4 \Big]\ln (u_2) \Bigg\}\nonumber\\[7pt]
\hat{\Gamma}_{S^P}^+ &=& (1+\epsilon_b)\Bigg\{ \zeta^2
     \frac{1+y}{1-y}
     \Big[ \zeta^2-(1-y)^2 \Big] \nonumber \\ [1.5mm]
     && \hspace{1.5cm} \times \Big[ (1-y)^2
     \left(3+4y+3y^2\right)
     +\left(3-4y+3y^2\right) \zeta^2 \Big]
     \nonumber \\ [1.5mm]
     &&+\zeta^2 \left(1-y^2\right)
     \Big[ \left(1-y^2\right)^2
     +4 \left(1+y^2\right) \zeta^2+\zeta^4 \Big]
     \ln (u_3) \Bigg\} \nonumber \\ [1.5mm]
     &-& K_b\Bigg\{ \frac{1}{3}\zeta^2
     \frac{1+y}{1-y}\Big[
     \zeta^2-(1-y)^2 \Big] \nonumber \\ [1.5mm]
     && \hspace{1.5cm} \times \Big[ (1-y)^2
     \left(1+12y+9y^2
    \right)+\left(1-12y
     +9y^2\right) \zeta^2 \Big] \nonumber \\ [1.5mm]
     &&+\frac{1}{3}\zeta^2 \left(1-y^2\right)
     \Big[ 3 \left(1-y^2\right)^2
     -4 \left(1-3y^2\right) \zeta^2+3\zeta^4 \Big]
     \ln (u_3) \Bigg\}\nonumber\\[7pt]
\hat{\Gamma}_{SL}^+ &=& \zeta^2\frac{1+y}{1-y}
     \Big[ \zeta^2-(1-y)^2 \Big]\Big[ (1-y)^2
     \left(3+4y+3y^2\right)
     +\left(3-4y+3y^2\right) \zeta^2 \Big]
     \nonumber \\ [1.5mm]
     &&+\zeta^2 \left(1-y^2\right)
     \Big[ \left(1-y^2\right)^2
     +4 \left(1+y^2\right) \zeta^2+\zeta^4 \Big]\ln (u_3)
     \nonumber \\ [1.5mm]
     &+& K_b\Bigg\{-\frac{1}{3}\zeta^2
     \frac{1+y}{1-y}
     \Big[ \zeta^2-(1-y)^2 \Big] \nonumber \\ [1.5mm]
     && \hspace{1.5cm} \times \Big[ (1-y)^2
     \left(1+12y+9y^2\right)
     +\left(1-12y+9y^2\right) \zeta^2 \Big]
     \nonumber \\ [1.5mm]
     &&-\frac{1}{3}\zeta^2 \left(1 -y^2\right)
     \Big[ 3 \left(1-y^2\right)^2
     -4 \left(1-3y^2\right) \zeta^2+3\zeta^4
     \Big]\ln (u_3) \Bigg\} \nonumber \\ [1.5mm]
     &+& G_b\Bigg\{-\, \frac{1}{3}\zeta^2
     \frac{1-5y^2}{(1-y)^2}
     \Big[ \zeta^2-(1-y)^2 \Big] \nonumber \\ [1.5mm]
     && \hspace{1.5cm} \times \Big[ (1-y)^2
     \left(1+12y+9y^2\right)
     +\left(1-12y+9y^2\right) \zeta^2 \Big]
     \nonumber \\ [1.5mm]
     &&-\frac{1}{3}\zeta^2 \left(1-5y^2\right) \Big[
     3 \left(1-y^2\right)^2
     -4 \left(1-3y^2\right) \zeta^2
     +3\zeta^4 \Big]\ln (u_3) \Bigg\}\nonumber\\[7pt]
\hat{\Gamma}_{SL^P}^+ &=&-\, (1+\epsilon_b)\Bigg\{
     \sqrt{R}\zeta^2\Big[ 3-4y^2+3y^4+3
     \left(1+y^2\right) \zeta^2 \Big] \nonumber \\ [1.5mm]
     &&+2\frac{y^4\zeta^2}{1-y^2}
     \Big[ \left(1-y^2\right) \left(3-y^2\right)
     -4 \left(1-y^2\right) \zeta^2+\zeta^4 \Big]\ln (u_1)
     \nonumber \\ [1.5mm]
     &&+2\frac{\zeta^2}{1-y^2}
     \Big[ \left(1-y^2\right)^4
     +4 \left(1-y^2\right)^2 \left(1+y^2\right) \zeta^2
     +\left(1+y^4\right) \zeta^4 \Big]\ln (u_2) \Bigg\}
     \nonumber \\ [1.5mm]
     &+& K_b\Bigg\{ \frac{1}{3}\sqrt{R}\zeta^2\Big[
     5-16y^2+9y^4
     -\left(11-9y^2\right) \zeta^2 \Big] \nonumber \\ [1.5mm]
     &&-\frac{2}{3}\,\frac{y^2\zeta^2}{1-y^2}
     \Big[ \left(1-y^2\right)
     \left(8-9y^2+3y^4\right) \nonumber \\ [1.5mm]
     && \hspace{1.5cm}-4 \left(4-7y^2+3
     y^4\right) \zeta^2+\left(8-3y^2\right)
     \zeta^4 \Big]\ln (u_1 \nonumber \\ [1.5mm]
     &&+\frac{2}{3}\,\frac{\zeta^2}{1-y^2}
     \Big[ 3 \left(1-y^2\right)^4
     -4 \left(1-y^2\right)^2
     \left(1-3y^2\right) \zeta^2 \nonumber \\ [1.5mm]
     && \hspace{1.5cm}-\left(5+8y^2-3
     y^4\right) \zeta^4 \Big]\ln (u_2) \Bigg\}\nonumber\\[7pt]
\hat{\Gamma}_{ST^P}^+ &=& (1+\epsilon_b)\Bigg\{ 
     \frac{4\sqrt{2}}{15}\sqrt{R}\,\frac{\zeta^3}{1-y^2}
     \Big[ \left(1-y^2\right) \left(8-7y^2\right)
     +\left(8-3y^2\right) \zeta^2 \Big] \nonumber \\ [1.5mm]
     &&-\frac{4\sqrt{2}}{15}\,\frac{\zeta^2}{1-y}\Big[
     \left(1-y^2\right)
     \left(2-7y^2-3y^4\right) \nonumber \\ [1.5mm]
     && \hspace{1.5cm}+10 \left(2-3y^2+y^4\right)
     \zeta^2+5 \left(2-y^2\right) \zeta^4 \Big]
     \Big[ \mbox{E} \! \left(k^2\right)-\mbox{E} \!
     \left(\varphi, k^2\right) \Big] \nonumber \\ [1.5mm]
     &&+\frac{8\sqrt{2}}{15}
     \frac{y\zeta^2}{1-y^2}\Big[
     \left(1-y^2\right) (1+y)
     \left(2-3y-3y^2\right) \nonumber \\ [1.5mm]
     && \hspace{1.5cm}+10 \left(1-y^2\right)
     (2-y)\zeta^2+5 \left(2-y\right) \zeta^4 \Big]
     \Big[ \mbox{K} \! \left(k^2\right)-\mbox{F} \!
     \left(\varphi, k^2\right) \Big] \Bigg\} \nonumber \\ [1.5mm]
     &+& K_b\Bigg\{ \frac{8\sqrt{2}}{45}\sqrt{R}
     \frac{\zeta^3}{1-y^2}\Big[ \left(1-y^2\right)
     \left(7-8y^2\right)-\left(3-8y^2\right)
     \zeta^2 \Big] \nonumber \\ [1.5mm]
     &&-\frac{8\sqrt{2}}{45}\,\frac{\zeta^2}{1-y}\Big[
     \left(1-y^2\right) \left(3+7y^2
     -2y^4\right) \nonumber \\ [1.5mm]
     && \hspace{1.5cm}+10 \left(1-y^2\right)
     \left(1-2y^2\right) \zeta^2-5 \left(1-2
     y^2\right) \zeta^4 \Big]\Big[ \mbox{E} \! \left(k^2\right)
     -\mbox{E} \! \left(\varphi, k^2\right) \Big] \nonumber \\ [1.5mm]
     &&+\frac{16\sqrt{2}}{45}
     \frac{y\zeta^2}{1-y^2}\Big[
     \left(1-y^2\right) (1+y)
     \left(3+3y-2y^2\right) \nonumber \\ [1.5mm]
     && \hspace{1.5cm}+10 \left(1-y^2\right)
     \left(1-2y\right) \zeta^2-5 \left(1-2
     y\right) \zeta^4 \Big]\Big[ \mbox{K} \!
     \left(k^2\right)-\mbox{F} \! \left(\varphi, k^2\right) \Big] \Bigg\}
    \nonumber\\[7pt]
\hat{\Gamma}_{I^P}^+ &=& (1+\epsilon_b)
     \frac{8\sqrt{2}}{15}\,\frac{\zeta^2}{1-y}
     (1-y-\zeta)^4\Big[ 1+3y
     +y^2-(1-y)\zeta \Big] \nonumber \\ [1.5mm]
     &-& K_b\frac{8\sqrt{2}}{45}
     \frac{\zeta^2}{1-y}(1-y-\zeta)^3
     \nonumber \\ [1.5mm]
     && \hspace{1.5cm} \times \Big[ (1-y)
     \left(3-6y-2y^2\right)
     +\left(19+2y+4y^2\right)
     \zeta-2(1-y)\zeta^2 \Big] \nonumber
\end{eqnarray}
(for $K_b$, $\epsilon_b$ and $G_b$ cf.\ again
Eqs.~(\ref{defKb})--(\ref{defGb})). The function $R$ is defined by
\begin{equation}
R = \lambda (1, y^2, \zeta^2) = 1+y^4+\zeta^4
     -2 \left(y^2+\zeta^2+y^2\zeta^2\right).
\end{equation}
In the analytical expressions for the polarized rates $\hat{\Gamma}_{I^P}^-$
and $\hat{\Gamma}_{ST^P}^+$ occur the elliptical integrals $\mbox{E}$,
$\mbox{F}$ und $\mbox{K}$ which are defined by
\begin{eqnarray}
\mbox{E} \! \left(\varphi, k^2\right) &=& \int_0^{\varphi}
     \sqrt{1-k^2\sin^2 t}d t, \hspace{2cm}
\mbox{E} \! \left(k^2\right) = \mbox{E} \! \left(\frac{\pi}{2}, k^2
    \right)\nonumber\\[7pt]
\mbox{F} \! \left(\varphi, k^2\right) &=& \int_0^{\varphi}
     \frac{d t}{\sqrt{1-k^2\sin^2 t}}, \phantom{d t} \hspace{2cm}
\mbox{K} \! \left(k^2\right) = \mbox{F} \! \left(\frac{\pi}{2}, k^2
    \right).
\end{eqnarray}
The argument $\varphi$ and the parameter $k$ of the elliptical integrals read
\begin{equation}
k = \frac{1-y}{1+y},\qquad
\varphi = \arcsin \left(\frac{\zeta}{1-y}\right).
\end{equation}

\end{appendix}

\end{document}